\documentclass[12pt,preprint]{emulateapj}

\def\kms{~km~s$^{-1}$\ }

\def\arcs{\char'175\ }
\def\arcsec{\char'175 }
\def\arcm{\char'023\ }

\def\hub{\ifmmode H_\circ\else H$_\circ$\fi}

\shorttitle{Star Clusters in M31}
\shortauthors{Caldwell et al.}

\begin{document}

\title{Star Clusters in M31: I. A Catalog and a Study of the Young Clusters }

\author{Nelson Caldwell} 
\affil{Smithsonian Astrophysical Observatory, 60 Garden Street, Cambridge, MA 02138, USA
\\ electronic mail: caldwell@cfa.harvard.edu}

\author{Paul Harding}
\affil{Department of Astronomy,
Case Western Reserve University, Cleveland OH 44106-7215
\\ electronic mail: paul.harding@case.edu}

\author{Heather Morrison}
\affil{Department of Astronomy,
Case Western Reserve University, Cleveland OH 44106-7215
\\ electronic mail: heather@vegemite.case.edu}

\author{James A. Rose}
\affil{Department of Physics and Astronomy, University of North Carolina,
  Chapel Hill, NC 27599, USA \\ electronic mail:
 jim@physics.unc.edu}

\author{Ricardo Schiavon}
\affil{Gemini Observatory, 670 N. A'ohoku Place , Hilo, HI 96720, USA \\ electronic mail:
rschiavo@gemini.edu}

\author{Jeff Kriessler}
\affil{Department of Astronomy,
Case Western Reserve University, Cleveland OH 44106-7215
\\ electronic mail: jeffrey.kriessler@case.edu}


\begin{abstract}

We present an updated catalog of 1300 objects in the field of M31, 
including 670 likely star clusters of various types, the rest being
stars or background galaxies once thought to be clusters. 
The coordinates in the catalog are accurate to 0.2\arcsec,  and are based
on images from the Local Group Survey \citep[LGS,][]{massey} or from the DSS.
Archival HST images and the LGS were inspected to confirm cluster
classifications where possible, but most of the classifications are
based on spectra taken of  $\sim1000$  objects with the Hectospec fiber positioner and
spectrograph on the 6.5m MMT.
The spectra and images
of young clusters are analyzed in detail in this paper; analysis of older
clusters will appear in a later paper.  Ages and reddenings of 140 
young clusters are derived by comparing the observed spectra with model spectra.
Seven of these clusters also have ages derived from HST color-magnitude diagrams (two
of which we present here); these agree well with the spectroscopically determined
ages.  Combining new V band photometry with the M/L values that correspond to
the derived cluster ages, we derive masses for the young clusters, finding
them to have masses as great as $10^5$ with 
a median of $10^4{\rm M}_{\sun}$, and a median age of 0.25 Gyr.
In comparison therefore, Milky Way open clusters have the lowest median mass, the
Milky Way and M31 globulars the highest, and the LMC young massive clusters
and the M31 young clusters are in between.
The young clusters in M31 show a range of structure. Most
have the low concentration typical of Milky Way open clusters, but
there are a few which have high concentrations.
We expect that most of these young clusters will be
disrupted in the next Gyr or so, however, some of the more massive and
concentrated of the young clusters will likely survive for longer.
The spatial distribution of the
young clusters is well correlated with the star-forming regions as mapped
out by mid-IR emission. A kinematic analysis likewise confirms the 
spatial association of the young
clusters with the star forming young disk in M31.

\end{abstract}


\section{Introduction}

In the Milky Way, there is a clear separation between known open clusters
(which have diffuse structure, generally have low masses and ages, and
belong to the disk) and globular clusters (which have a more
concentrated structure, higher masses and ages, and in many cases
belong to the halo). When the only well-studied
globular cluster system was that of the Milky Way, 
it was generally thought that this separation
was because globular clusters were fundamentally different from other
star clusters, perhaps because of conditions in the early universe
\citep{peebles68,fallrees85}.

However, it is possible to produce this apparent bimodality from
clusters formed in a single process, with the same cluster initial
mass function. In this picture, cluster disruption mechanisms, which
are more effective at destroying low-mass clusters in particular
because of two-body relaxation \citep{spitzer58,spitzerharm}, would remove almost
all of the low-mass older clusters.  If all clusters were born with
similar cluster mass functions, then we would expect to see the
occasional high-mass young cluster. In fact, we do see these in other
galaxies. The ``populous blue clusters'' of the LMC
\citep{kcf80,hodge81,mateo93} have been suggested as examples of young
objects which will evolve into globular clusters. M33 also has a few
likely massive young clusters \citep{chandar99} , and such clusters
have been found in a number of normal isolated spirals
\citep{larsen04}. It is possible that the seeming absence of such
objects in the Milky Way is merely an observational selection effect;
recently, there have been discoveries of  heavily
reddened open clusters such as Westerlund 1, which likely has mass in
excess of $10^5 $M$_\odot$ \citep{clark}.

What of M31's clusters? While its clusters have been studied since the
1960s \citep[eg][]{kinman,vetesnik,vdb} and it was noted even then that
some of these clusters had colors indicating young populations, 
their nature is still not entirely clear. \citet{vdb} called them open
clusters, while \citet{krienke} adopted the simple convention of
calling any cluster projected on M31's disk a ``disk cluster'' until
proved otherwise by kinematics. This avoids the question of whether
young clusters are fundamentally different from globular clusters in
structure, formation, etc. Our detailed study of M31 young clusters,
incorporating kinematics, should cast some more light on these
questions.

It is only recently that detailed constraints on the mass, kinematics,
age and structure of cluster populations in M31 have been obtained,
particularly for clusters projected on the inner disk and
bulge. Multi-fiber spectroscopy and HST imaging have played an
important role here.  A number of M31 globular cluster catalogs have
been created over the years, giving a very heterogeneous result, with
significant contamination by both background galaxies, foreground
objects and even non-clusters in M31 itself. While the work of
\citet{perrett}, \citet{barmby}, \citet{galleti2}, \citet{kim}, and \citet{lee} has gone a long way towards
cleaning up the catalogs and winnowing out the non-clusters, still
more work is needed for both young and old clusters.  Here we present
a new catalog of M31 star clusters which were originally classified as
globular clusters, all with updated high-quality coordinates. We have
observed a large number of these clusters with the MMT and the
Hectospec multi-fiber system. In this paper we study more than 100
young M31 clusters in detail.  In subsequent papers we will address
the kinematics, ages and metallicities of the older clusters.

The M31 young clusters have been studied both by authors aiming to study
its open clusters \citep[eg][]{hodge79,hodge87,williams01,krienke},
and also by others who have aimed to study its globular clusters.  For
example, \citet{elsonw} pointed out the existence of young clusters in
M31, estimated masses between $10^4$ and $10^5 M_\odot$, and drew
attention to their similarity to the populous blue clusters in the
LMC. \citet{barmby} noted the existence of 8 clusters with strong
Balmer lines in their spectra, which they tentatively classified as
young globular clusters. \citet{beasley} \citep[confirmed with a later
sample by][]{puzia} added more clusters, and commented that HST
imaging (then unavailable) was needed to distinguish between structure
typical of open and globular clusters. \citet{burstein} added more new
young clusters, bringing the total to 19, and \citet{fusipecci}
increased the sample to 67. In general, authors have associated these
young clusters with M31's disk, although \citet{burstein} invoke an
accretion of an LMC-sized galaxy by M31.  

Observations are particularly challenging for clusters projected on
M31's disk: many of the early velocities had large errors, and there
were issues with background subtraction. Here we discuss high-quality
spectroscopic measurements of kinematics and age for the young
clusters, supplemented with HST imaging to delineate the structural,
spatial and kinematical properties of these young clusters.  We find
that while they are almost all kinematically associated with M31's
young disk, and their age distribution will allow us to test
suggestions that M32 has had a recent passage through M31's disk
\citep{gordon,block}.
The young  clusters have structure and masses which range all the way
from the low mass Milky Way open clusters to higher mass, more
concentrated globular clusters, although they are dominated by the
lower concentration clusters. We will also discuss the likelihood that
these clusters will survive.

\section{Revised Catalog of M31 Clusters}

The starting point for our cluster catalog was the Revised Bologna
Catalog (RBC) \citep{galleti1, galleti2}, itself a compilation of
many previous catalogs.    Coordinates from this catalog were used to
inspect images from the Local Group Survey (LGS)
images \citep{massey}, which cover a region out to 18 kpc radius on the
major axis and 2.8 kpc away from the major axis, HST archival WFPC2 and 
ACS images, and the DSS for the outermost
clusters. The LGS images have median
seeing $\sim$ 1 \arcs.  We also added in some new clusters, 
found visually by ourselves on the LGS images and 
on HST images (discussed below).  Even a casual inspection of the LGS
images, particularly the I band images,  reveals the presence of
a vast number of uncataloged faint disk clusters, presumably similar to the
galactic open clusters \cite[estimate 10000 such clusters from
HST images]{krienke} .  We have elected not to take on the enormous
task of cataloging and measuring those clusters in this paper; rather we
choose to deal with the more massive clusters for which some
information, however fragmentary, already exists.
At a later stage in the project, the catalogs
of  \citet{kim}, extracted from KPNO 0.9m images, and \citet{krienke}, from 
HST images  became available, from which we collected
objects not already in our own catalog and subjected them to
the same editing as we now describe.  The archival images were
thus used to answer the following questions from the catalog we created
from the RBC and our own additions.

First, did the catalog coordinates correspond to a unique object?
In cases where the identification of the cluster
on the Local Group Survey was ambiguous or unclear, we consulted the
original papers and finding charts where these were available
\citep{vetesnik,baade_arp,sargent,battistini80,battistini87,battistini93,crampton85,racine,auriere,poles,barmby_huchra}.
In some cases, there is no clear object that can be associated with the published coordinates, or
the nearest object in fact already had a different ID. 
The
large number of Hectospec fibers meant that we were able to verify
classifications of many low-probability candidates.  The cataloging
and observation parts of this program occurred in a feedback
fashion, allowing some target names and/or coordinates to be changed for
the succeeding observations. As a result, we had some objects
whose identifications were incorrect; to these we add an ``x'' to
their original name in our tables below.  

Second, were the existing coordinates accurate? In general, the answer
was no.  Our final catalog contains 1200 objects from the
RBC, without considering the newer candidates of 
\citet{kim} and \citet{krienke}. 830 of those required coordinate corrections larger
than 0.5\arcs to place them on the FK5 system used in the DSS and the LGS
images.  The median error in coordinates is 0.8\arcs, with the
largest error being of order 10\arcs; at which point the identification
of the actual object becomes uncertain. Similarly, there are
379 objects in the \citet{kim} catalog found within 2\arcs of an LGS object. For these, the median error is
0.8\arcs offset, where the largest error is 1.9\arcs.  Many of
the discrepant velocities between us and the RBC or \citet{kim} tabulations
reported here are likely due to 
inaccurate coordinates used in previous spectroscopic work.  
The coordinates newly derived from the LGS images are accurate to 0.2\arcs \ 
rms.

Third, were the targets really clusters? The LGS V and I band images, and WFPC2 or ACS images taken with non-UV filters were used to confirm
the cluster nature of the objects, by visual inspection as well as the
automated image classifier contained in
the SExtractor code \citep{1996A&AS..117..393B}.  A number
of cluster candidates were stellar on the LGS images; all of these were
later confirmed as stars in our spectra, from either the spectral characteristics
or the velocities, in regions of M31 where there is no confusion between the local M31 velocity
field and the velocity distribution of foreground galactic stars. 
We found that about 90 of the \citet{kim} candidates listed as new, probable
and possible (indicating that they appeared non-stellar in their KPNO
0.9m images) appeared  stellar on the LGS or HST images.  
Objects for which we have no new classification data are kept in the catalog,
but noted as still in question in our table.

The large majority of the misclassified objects are
stars (foreground galactic or M31 members); more than 130 objects
considered to be clusters as recently as 2007 by \cite{galleti2} are in fact stars.
Quite a number of cataloged objects are background galaxies,  and a few are either unidentifiable,
or are accidental clumpings of galactic or M31 member stars.

\begin{figure*}[ht]
\includegraphics[scale=1.0,clip=true]{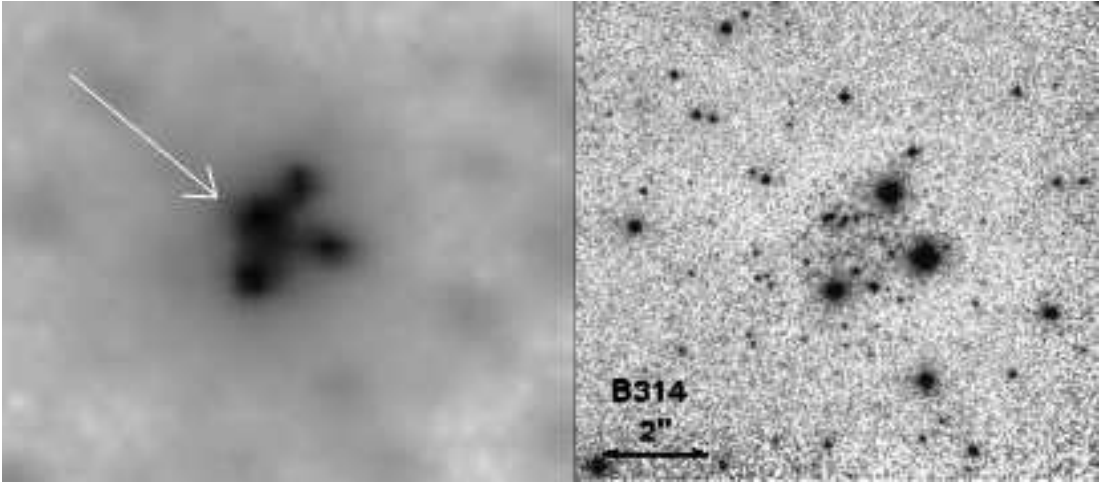}
\caption{The disputed cluster B314-G037. The LGS I band image is shown
on the left, next to the \citet{cohen} AO image, taken in the K\arcm band.
The I band reveals the star cluster clearly (arrow), though the magnitude
measured for the cluster previously using a large aperture was certainly
too bright. \label{cohen}}
\end{figure*}

\begin{figure*}[ht]
\centering
\includegraphics[scale=1.0,clip=true]{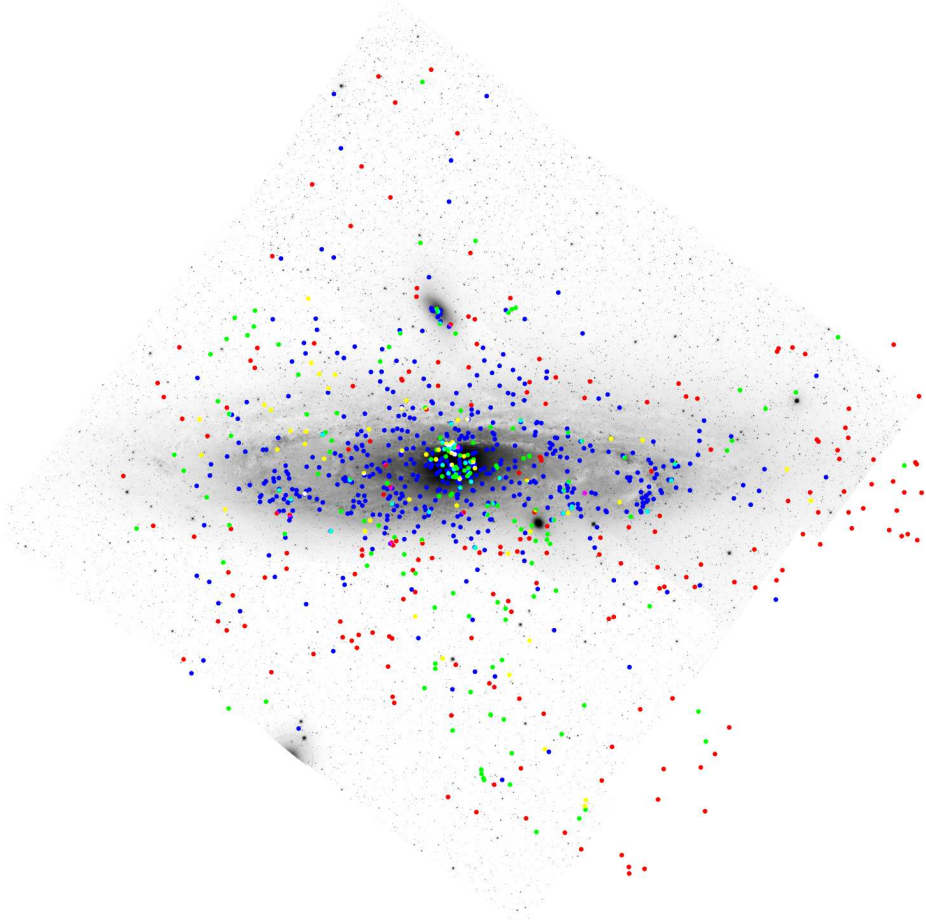}
\caption{Locations of the spectra taken with Hectospec 
of M31 cluster candidates. Confirmed clusters, stars, possible stars
and background galaxies are shown in blue, green, yellow and red, 
respectively.  \label{hectospec}}
\end{figure*}

\begin{figure*}[ht]
\centering
\includegraphics[scale=0.7,clip=true]{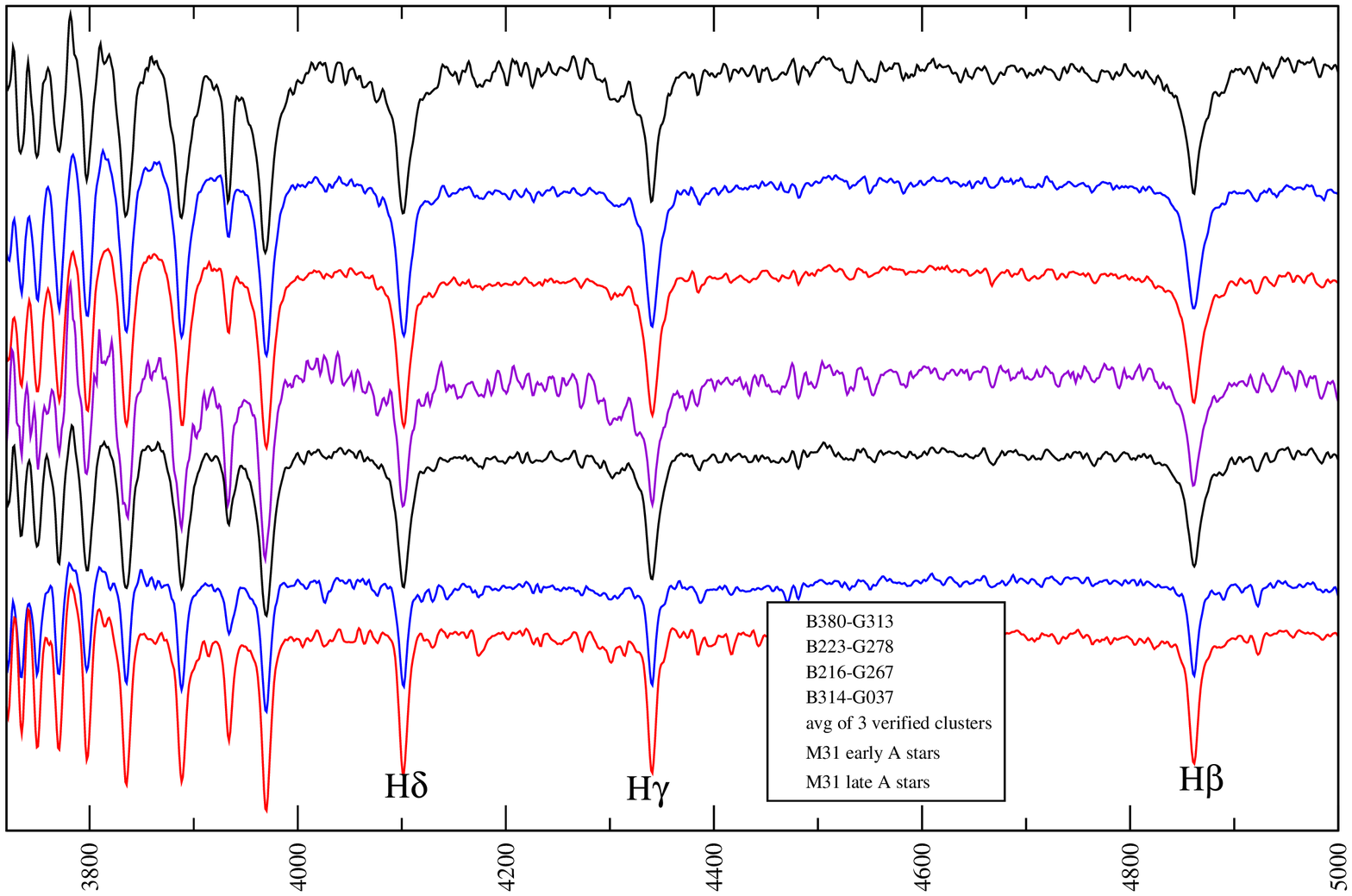}
\caption{Hectospectra of young clusters in M31. Shown are spectra of
clusters disputed as real by
\citet{cohen}, the average of three young clusters verified by ACS images, and the average
of single, blue supergiant stars as a comparison.  If the disputed
clusters were in fact merely a few stars mistaken for a cluster, their absorption line widths,
particularly the Balmer lines,
would be narrow as seen in the blue supergiant spectra.  These spectra have been gaussian smoothed and have had their
continuum shapes removed for ease of display.
\label{blue_spectra}}
\end{figure*}

\citet{cohen} have recently stated that four  clusters
that were classified as disk clusters by \citet{diskglobs} are 
``asterisms''. They note that in their Adaptive Optics (AO) images there was no
cluster visible -- generally there were only a few bright
stars. However, for young clusters, red supergiants would dominate the
light at infrared wavelengths and the hotter mainsequence stars would
appear much fainter. We would need to use multi-wavelength data to classify
these objects correctly. 
We show below that the optical spectra of those four clusters are indeed  consistent with clusters of massive, main-sequence stars,
and although the
magnitudes, and hence masses,  of these few objects were certainly overestimated, the objects
will still be considered as clusters in our catalog, at least until HST images
show otherwise.
Figure \ref{cohen} shows this 
complication for one cluster, by comparing the high resolution, but 
long wavelength AO image with the LGS I band image. A star cluster is clearly
visible in the I band, and even more prominent at bluer wavelengths - the
object is indeed a young star cluster, though certainly not a globular and not
very massive.
Our own HST images do reveal two cataloged clusters as asterisms:
these are comprised of a small number of OB stars
and late supergiants, resulting in a distinctive integrated spectrum
with strong Balmer and He I lines in the blue, and TiO bands in the
red. Even if  these two are real clusters, the derived
masses are small enough to exclude them from a list of massive clusters. 

Some cataloged objects have no real object even within a generous
radius.  In a few cases, a nearby background galaxy had also led to
confusion in previous papers (though not in the most recent
version of the RBC), whereby an actual cluster
was labeled as background. Thus, while for the most part we have removed
objects from the list of clusters, we have also restored a few objects to
the cluster list.

Table \ref{main} lists all objects believed to be clusters.  
For object names, we use the naming convention of \cite{barmby} where possible, where
the name consists of a prefix with the RBC number followed by the number of the
object from the next most significant catalog.  
Objects for which we have no new information 
other than improved coordinates, and which have not been convincingly
shown to be clusters by previous workers, are italicized.
Objects in the RBC which we did not observe and 
for which our coordinates  are within 0.5\arcs
of the RBC coordinates are not listed here, nor are the \citet {williams01,kim} or 
\citet{krienke} cluster
candidates that we did not observe.
Some objects that we did not observe could of course still be background
even though they have non-stellar profiles, but these, few in number, 
are still listed here.

A rough classification based on the spectra is included in this table, for
objects with good quality spectra. ``Young'' clusters are those
with ages less than 1 Gyr, ``interm'' refers to those with ages between 1 and
2 Gyr, and ``old'' refers to clusters older than that.  A subsequent paper
will provide a detailed analysis and evidence that few if any of these ``old'' clusters
have ages less than 10 Gyr. ``HII'' indicates the spectrum is emission-line dominated. ``na'' appears for objects known to be clusters from an HST image, but for which we have no spectrum, or cases where the spectrum is too poor to determine the age, even in a coarse manner.
The V magnitudes come from this paper, using the aperture size listed (see \ref{s:phot}) 
or otherwise as indicated. Column C describes what information was used to classify the
target as a cluster.  The possibilities are ``S'', where our spectrum clearly indicates
a star cluster, ``L'', where the LGS image is non-stellar, and/or ``H'', where an HST
image indicates a star cluster.

Table \ref{young} gives a list of those clusters that have ages less
than 2 Gyr. (Sections \ref{s:acs}, \ref{s:ages} and \ref{s:masses} of
this paper will discuss the measurement of ages and masses for these clusters.)
Table \ref{stars} lists objects from
previous cluster catalogs that are in fact stars.  Many of these had
also been classified as stars by previous workers. Asterisms are also listed
here.  Table
\ref{maybestars} gives a list of possible stars.  In these cases, the
Local Group Survey imaging indicates a stellar profile, but either we
have no spectrum, or the spectrum is ambiguous.  Note that some of
the stars in Tables \ref{stars} and \ref{maybestars} are certainly members 
of M31. Objects thought to be clusters
in the \citet{kim} catalog but which have stellar profiles in the LGS images
are listed in  \ref{maybestars}, with coordinates derived from the LGS.  Table \ref{gals} lists
background galaxies. Table \ref{nobodyhome} lists cataloged objects
where there was no obvious object within a reasonable distance of the
previously published coordinates, which are listed here again.

As a commentary on the difficulty experienced by all of those who have 
endeavored to collect true M31 star clusters  (including us), here is a brief summary of
the contents of the most recent version of the RBC, excluding the additions of \citet{kim,williams01,huxor} and 
\citet{krienke}, but including the lists compiled by other astronomers
starting with Edwin Hubble.  The RBC, restricted as just mentioned, contains
1170 entries. We here, and others  \citep[particularly][]{barmby},  have provided classifications
for 991 of these. Only 620 entries are actually star clusters, while 20 more 
could be considered clusters though the large amount of ionized gas present indicates
the clusters may still be forming. 270 entries are stars, mostly foreground, and 224 entries
are background galaxies or AGN. At least 2 objects are chance arrangements of luminous
M31 stars, together which appear as clusters from the ground. 

In the  \citet{kim}
catalogs, there 113, 258 and 234 ``new'', ``probable'' and ``possible'' clusters, 
respectively.  The LGS survey contains images of 94, 152 and 105 members
of those catalogs, respectively.  Of those subsets, 79, 106, and 129, respectively
have non-stellar profiles in the LGS, some of which were observed with Hectospec.

\section{Hectospec Observations}

The Hectospec multi-fiber positioner and spectrograph is ideally suited
for this project, in that its usable field is 1 degree in diameter, and
the instrument itself is mounted on the 6.5m MMT telescope.
We obtained data in observing runs in the years
2004 to 2007, and  now have high-quality spectral observations of over
400 confirmed clusters in M31, and lower quality spectra of 50 more. 
We used the 270 gpm grating (except for a small number of objects
taken with a 600 gpm grating), which gave
spectral coverage from 3650--9200\AA \ and a resolution of $\sim$5\AA.
The normal operating procedure with Hectospec, and other multi-fiber
spectrographs, is to assign a number of fibers to blank sky areas in
the focal plane, and then combine those in some fashion to allow
sky subtraction  of the target spectra.  For instance, the 4-5 fibers
nearest on the sky to the target may be combined.
These methods are satisfactory for our outer M31 fields, but
not for the central areas where the local background is high relative
to the cluster targets.  For those fields,
we alternated exposures on-target and off-target to allow local
background subtraction to be performed. Many of the discrepancies between our
bulge cluster velocities and those of previous workers might be explained
by their lack of proper background subtraction, and/or inaccurate target coordinates.   

We obtained exposures for 25 fields,  with total exposure times varying
between 1800s and 4800s. The signal/noise ratios for the 500 objects we
classified as clusters have a median
of 60 at 5200\AA \ and 30 at 4000\AA, with more than 100 clusters having
a ratio at 5200\AA \ better than 100.
Figure \ref{hectospec} shows the locations and types of
objects observed in all of these fields.

The multifiber spectra were reduced in a uniform manner. For each field,
or configuration, the separate exposures were  
debiased and flat fielded, and then
compared before extraction to allow identification
and elimination of cosmic rays through interpolation.  Spectra were then 
extracted, combined and wavelength calibrated.  Each fiber has a  distinct
wavelength dependence in throughput, which can be estimated using
exposures of a continuum source or the twilight sky. The object spectra
were thus corrected for this dependence next, followed by a correction
to put all the spectra on the same exposure level. The latter correction was
estimated by the strength of several night sky emission lines.   Sky subtraction
was performed, using object-free spectra as near as possible to each target.
For the targets where local M31 background was high, the method was reversed,
such that only sky spectra
far from the disk of M31 were used.  An offset exposure for such fields, taken concurrently, was reduced in 
a similar way (thus contemporaneous sky subtraction was performed for on- and
off-target exposures), and then the off-target
spectra were  subtracted from the on-target.  This process  increased
the resultant noise of course, but we deemed it essential for targets in the bulge
and disk of M31.  The off-target exposures have the additional advantage of
giving measurements of
the unresolved light in over 800 locations over the entire disk of M31.

Velocities were measured using the SAO xcsao software. Given the wide variety
of spectra in this study, it was deemed necessary to develop new velocity
templates, from the spectra themselves.  The procedure was to derive an initial
velocity of all spectra using library templates (typically a K giant star). The
spectra were shifted to zero velocity, and sorted into three different spectral types, A, F
and G type spectra.  The best spectra in each group were combined to make new templates,
and the procedure was repeated now using the new templates.  By using these templates
we have assured that all the M31 targets are on the same velocity system.  They are tied
to an external velocity in the initial step, whose accuracy depended on the accuracy of
the initial set of templates  used.  A good test of the internal accuracy was provided
by repeat measurements of clusters. We have 386 repeat measurements (on different nights)
for 224 clusters. The median difference in velocity for these repeats was 0.5 \kms, with
an implied median single measurement error of 11 \kms (smaller than our formal errors listed
in the tables).    We will present external comparisons in a subsequent paper,  but note that
the cluster velocities agree very well with the HI rotation curve (see \ref{s:kin}). Velocities for the young
clusters, stars and galaxies are presented in Tables \ref{young}, \ref{stars},  \ref{maybestars},
and \ref{gals}. Velocities for the old clusters will presented in a subsequent paper.

The spectra were corrected to relative flux values, using observations of
standard stars (the MMT F/5 optics system employs an atmospheric dispersion
compensator, ADC).  The flux correction
has been determined to be very stable over several years.  Thus, observations
of the same targets taken in different seasons can be combined where available.

\section{Using HST images to Determine Cluster Classification }
\label{s:acs}

\citet{cohen} highlighted the heterogeneous quality of the M31 cluster
catalogs when they claimed, using AO techniques at K\arcm, that four
out of six observed young clusters were in fact asterisms.  Figure
\ref{blue_spectra} shows, from top to bottom, spectra of the four
disputed clusters, the average of three genuine young clusters
verified by ACS images, and the average spectra of single supergiant
stars (these were verified to be stars from the LGS images, and members
of M31 from their velocities).  If the disputed clusters were in fact
merely a few stars, those stars would have to be supergiant stars,
whose absorption line widths would be as narrow as seen in our blue
supergiant spectra. This is not the case for the four disputed
clusters (note in particular the H$\beta$ and H$\delta$ widths, narrow
in the stars and wide in the other spectra),
and we conclude that those objects are true clusters and not
asterisms. To be sure, these particular clusters are not globular
clusters either, and, additionally, are perhaps not massive enough to
be considered young, populous clusters.

High spatial resolution imaging can both check for
asterisms and also explore the clusters' spatial structure: is their
concentration low, like typical Milky Way open clusters, or high, like
globular clusters?

\begin{figure*}[ht]
\includegraphics[scale=0.6,clip=true]{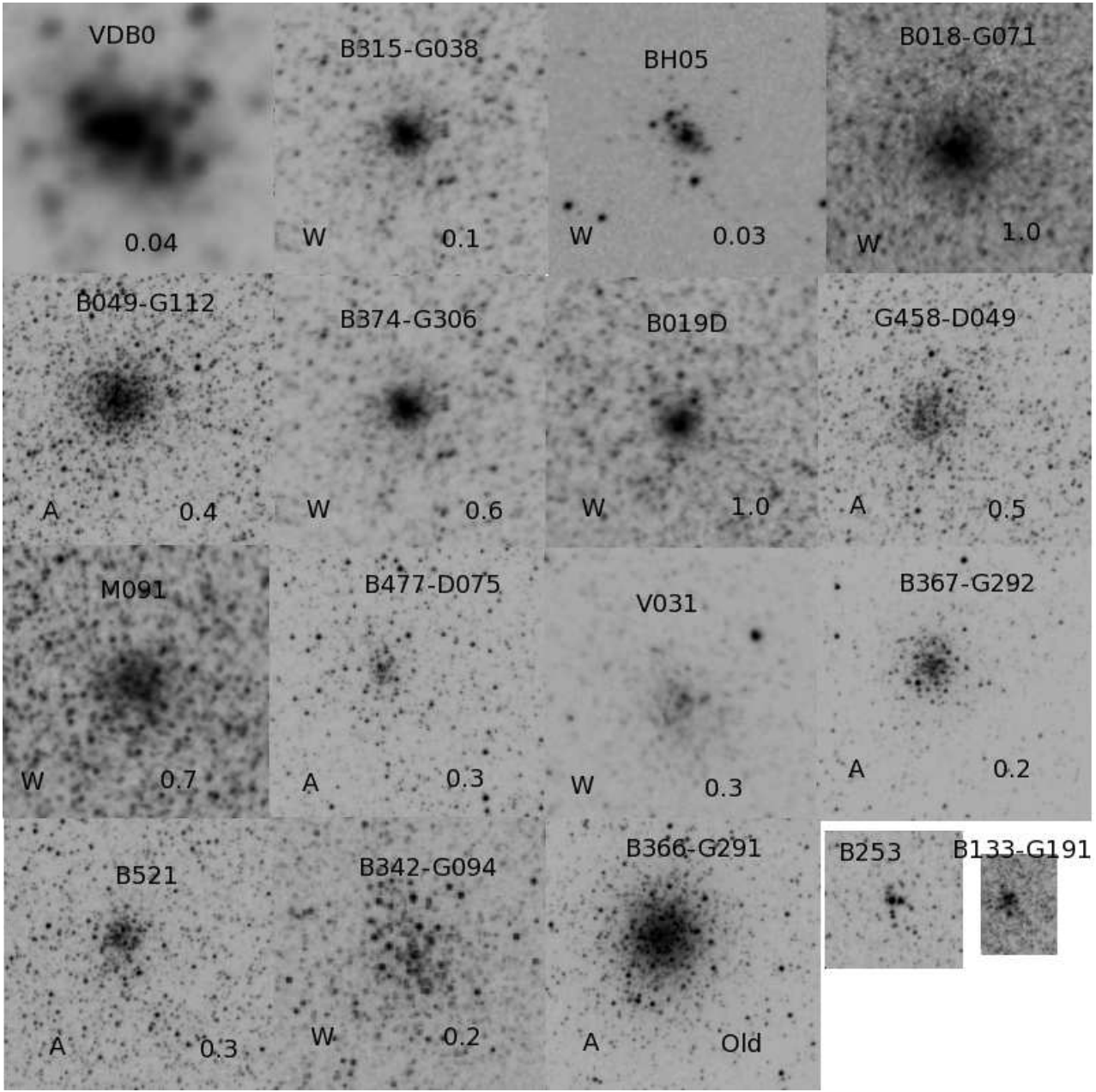}
\caption{HST ACS or WFPC2 images of a selection of clusters with ages less than 2 Gyr, except for
VDB0 whose image is from the LGS V band.  The HST images were taken in either F555W or F606W.
The spectroscopically  determined  ages in Gyr are listed for each image, as is the camera used (``A''=ACS, ``W''=WFPC2), to aid in comparison since in 
general the ACS images are deeper. A comparison old M31 globular cluster is shown, as are
two candidates that turned out to be asterisms.  The scale is the same for all images; except for the
two small ones, each image is 10\arcs on a side. The images are ordered by descending mass, starting at the top left. \label{acs_images}}
\end{figure*}

There are ACS or WFPC2 images available for 25 of the clusters with
ages less than 2 Gyr.  Two of these show no evidence of an underlying
cluster, but the remaining 23 are clearly not
asterisms. Figure \ref{acs_images} shows the range of structure seen
in these young clusters. While many of them show the low-concentration
structure typical of Milky Way open clusters, a number of them, such
as B374-G306 and B018-G071, are quite centrally concentrated, like the
majority of the Milky Way globular clusters.

\citet{barmbystruc} have measured the structure of M31 clusters with
available HST imaging at the time of publication. There are 70
clusters in their sample which we have classified as old, and 7 of our
young clusters. It is straightforward to compare the structure of the
clusters they study with the Milky Way globulars, because they use the
same technique as \citet{dean}, who have produced a careful summary of
the structure of the Milky Way globular clusters. However,
it should be noted that their fitting technique (fitting ellipses to
cluster isophotes) is not well-suited to very low-concentration
clusters, and in fact one of our young clusters, B081D, is omitted
from their analysis because of its low density. 

\begin{figure*}[ht] \includegraphics[scale=0.4,clip=true]{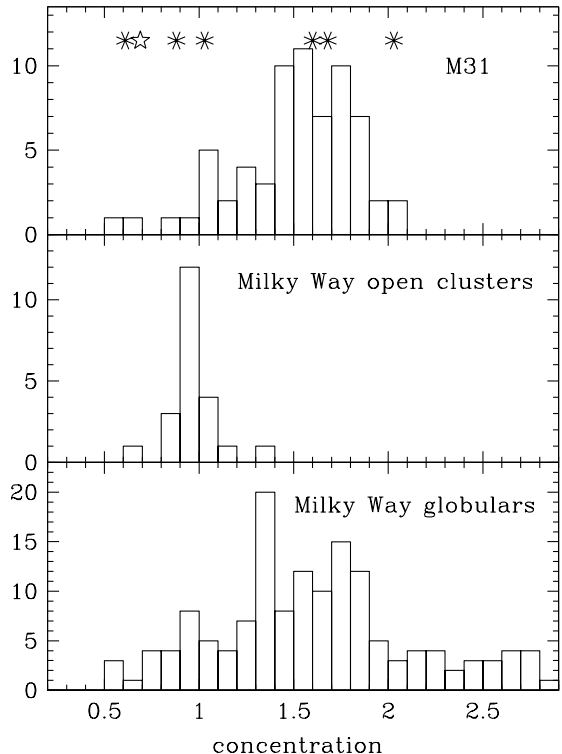}
\centering
\caption{Histograms of the concentration parameter from King model
  fits for (top panel) old M31 clusters (young clusters shown by
  asterisks, the young cluster in NGC 205 by a 5-pointed star), (middle
  panel) Milky Way open clusters and (bottom panel) Milky Way globular
  clusters. It can be seen that the M31 old clusters resemble the
  Milky Way globulars in their concentration, while the M31 young
  clusters cover the range of both open and globular
  clusters. \label{structure}} \end{figure*}

Figure \ref{structure} compares their results for old clusters from
our sample with the structure of Milky Way globulars \citep[from the
work of][]{dean}.  It can be seen that the concentration
parameter\footnote{$c$ = log($r_t/r_0$) for King model fits,
  \citet{bt} p. 307} for the
old clusters in M31 has a similar distribution to the Milky Way
globulars. 
We note that although there are no old clusters in 
our sample with concentration greater than 2.2, such 
clusters are definitely present in the \citet{barmbystruc} 
sample so this absence is unlikely to be significant.
The similarity in structure is interesting, because at first sight it would seem
that the M31 clusters with high concentration would be preferentially
discovered in surveys. Perhaps the M31 globular cluster surveys
(which, as we have seen above, include a large number of non-globular
clusters, as well as the low-concentration young clusters) are now
sufficiently thorough that they are not strongly affected by this bias.

Although only seven of our young M31 clusters were analyzed by
\citet{barmbystruc}, it can be seen from Figure \ref{structure} (where
they are marked by asterisks and five-pointed stars in the top panel)
that their concentrations cover the whole range of the older clusters
in M31 and in the Milky Way. (The five-pointed star represents
Hubble V from NGC 205.) Our observational selection biases may
over-emphasize high-concentration clusters, but it is still
interesting to see that three of the young clusters have quite high
concentration parameters. How does their structure compare with the
Milky Way open clusters? It is quite difficult to answer this question
because the available samples of Milky Way open clusters are severely
incomplete, and it is a challenging task to fit King models to the
known open clusters, because cluster membership is hard to determine.
The 2MASS database \citep{beichman98,jarrett} has been used by
\citet{bonatto05,santos,bonatto07,bica06,bica08} to measure the
structure of 21 open clusters. They used a CMD-fitting technique to
remove contamination from disk field stars. We have also used data
from \citet{eigenbrod}, who used radial velocity to decide membership.
Because of the high background in all these cases, it is possible that
the ``limiting radius" given by the authors is in fact smaller than the
tidal radius, in which case the cluster concentrations would be
smaller than those plotted. It can be seen in the middle panel of
Figure \ref{structure} that all these open clusters have quite low
concentrations.  However, the sample of clusters with concentration
measurements is quite small, and it is quite possible that there are a
few open clusters in the Milky Way which are yet to be discovered and
have high concentrations, like the two M31 young clusters.

In summary, the young clusters in M31 show a range of structure. Most
have the low concentration typical of Milky Way open clusters, but
there are a few which have high concentrations, like most Milky Way
globulars. We note that any survey of
M31 clusters will preferentially discover ones with high
concentrations. In addition, the incompleteness of Milky Way open
cluster samples and the difficulty of measuring cluster concentration
in crowded fields means that we cannot rule out the existence of such
clusters in the Milky Way.

\section{Ages of the Young clusters}
\label{s:ages}

In this section, we describe methods for determining ages from the spectra
and color magnitude diagrams for the verified clusters.   Since the
emphasis in this paper is on the younger clusters, and more specifically, on
their M/L ratios, our task is first 
to distinguish young from old clusters, and then to 
obtain accurate age measurements among the younger clusters.  A more refined age
determination (for clusters older than 2 Gyr) is postponed for a later
paper.

\subsection{Ages from Spectra}

\begin{figure*}[ht]
\centering
\includegraphics[scale=0.6,clip=true]{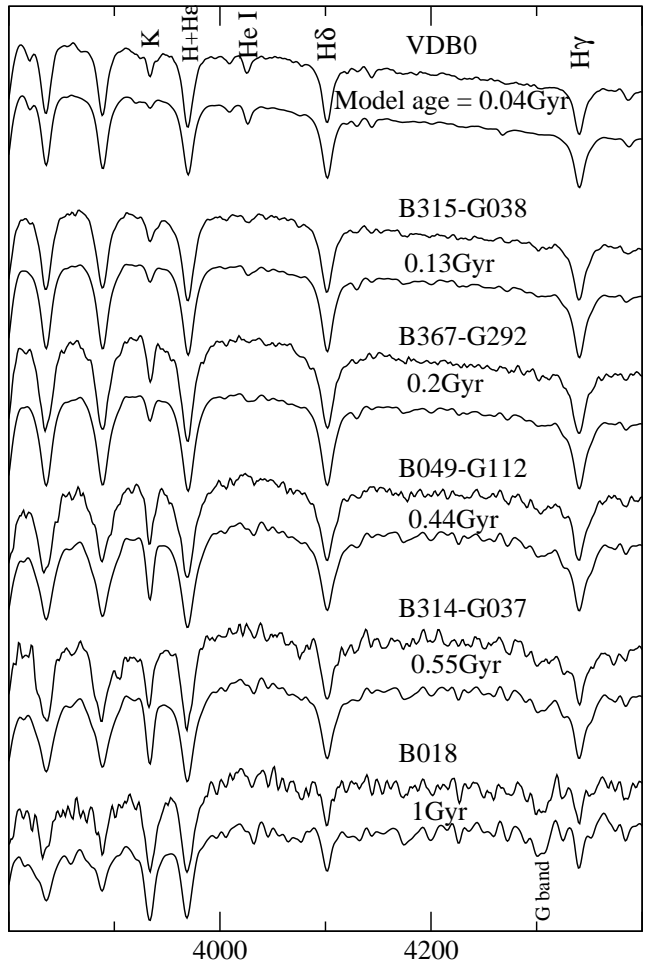}
\caption{Spectra of a sample of young clusters, ranging in ages from 0.04 to 1 Gyr.
Each object spectrum is shown with the best matching SB99 model spectrum.  The object
spectra have been smoothed with a gaussian with a sigma of 1.1\AA \, and have been
corrected for the reddening determined in this paper, which was itself determined by
matching the continuum shapes of objects and models.
\label{young_spectra}}
\end{figure*}

\begin{figure*}[ht]
\includegraphics[scale=0.6,clip=true]{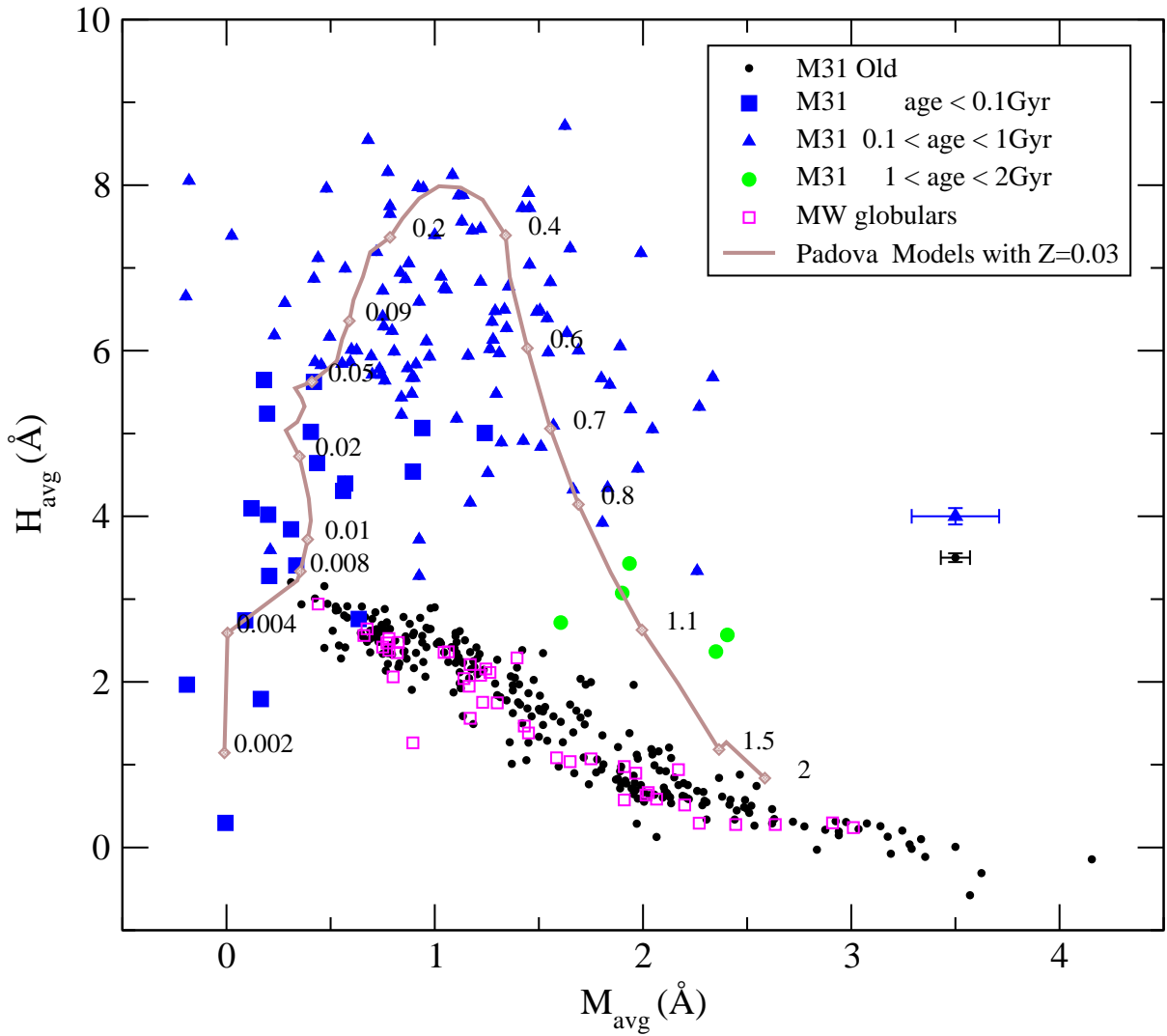}
\caption{M$_{\rm avg}$ vs H$_{\rm avg}$  for M31 cluster spectra, with different types identified. 
Model indices derived
from SB99 models using Padova Z=0.03 isochrones are also shown as a solid curve. Ages in Gyr are marked
at selected areas along the curve. Median error bars are shown at the right for young  and old
clusters separately. The maximum errors on points in this plot are 0.45 for M$_{\rm avg}$ and 0.19 for 
H$_{\rm avg}$.
\label{indices}}
\end{figure*}

\begin{figure*}[ht]
\includegraphics[scale=0.6,clip=true]{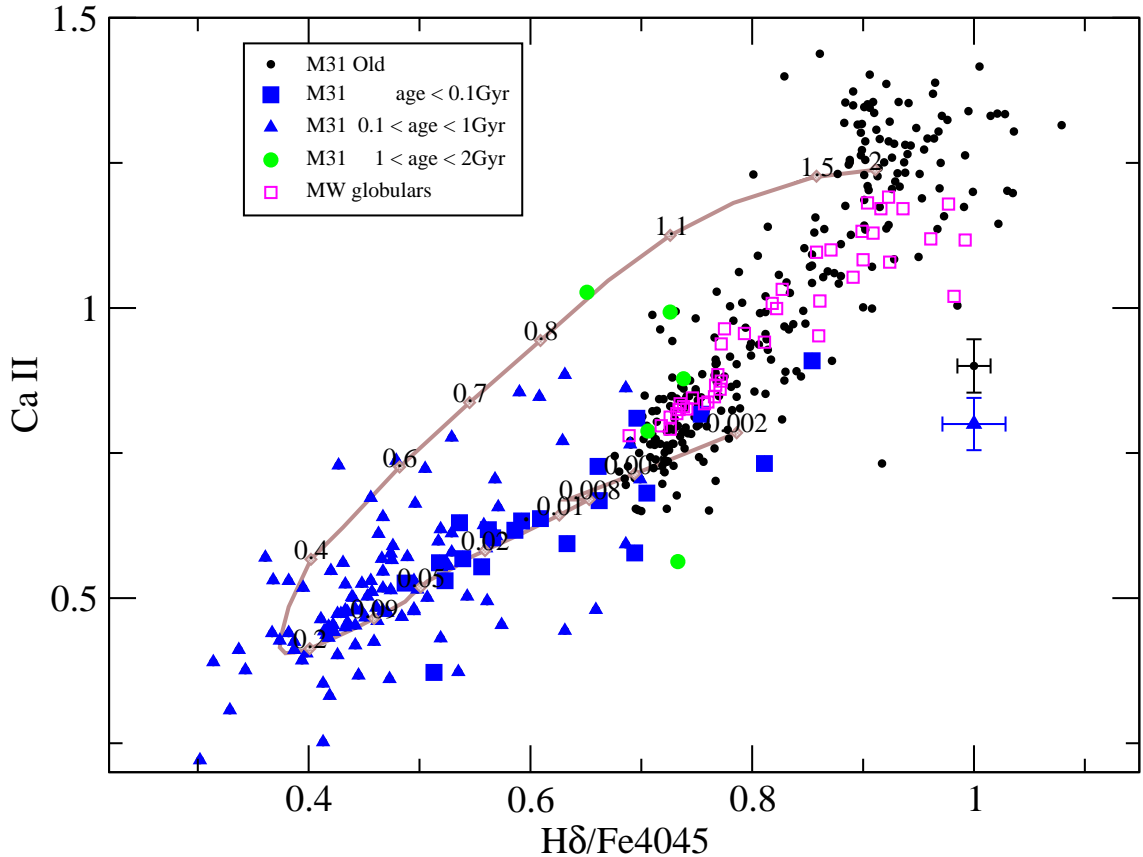}
\caption{Same as Figure \ref{indices} for H$\delta/$Fe4045 and CaII.  Maximum errors on 
points in this plot are 0.2 for both indices.
\label{indices2}}
\end{figure*}

The methods for obtaining ages for young stellar populations from their integrated 
spectra are similar to those 
used for older stellar populations, except that instead of employing 
empirical stellar libraries \citep[e.g.,][]{1994ApJS...95..107W},
modelers focussing on younger stellar
populations have used synthetic spectra,
partly due to a lack of empirical spectra of young stars over a range of metallicities.  
Here we have made use of the Starburst99 stellar population
modeling program \citep[SB99,][]{sb99}.

To distinguish young from old clusters, we compared our cluster
spectra with two external sets of spectra, which served as population
templates.  
One set was the sample of 41 MW globular spectra obtained by \cite{2005ApJS..160..163S},
covering the abundance range of $-2 < $[Fe/H]$ < 0 $. These spectra have
a wavelength coverage from 3500 to 6300\AA \ , a dispersion of 1\AA /pixel and
a resolution of about 4.5 \AA , and  served as our old population
templates.  Our young population templates were created from the 
SB99 program using  the Padova Z=0.05 interior models and Z=0.04
stellar atmospheres, since it seemed likely the young clusters have 
supersolar abundances.  Using a solar abundance set of isochrones in the models
has the net effect of increasing the derived ages for clusters older than 0.1 Gyr,
in a logarithmic scaling such that a 1 Gyr supersolar model and a 2.5 Gyr solar 
abundance model have nearly the same resultant spectra.
A grid of 40 high resolution spectra was created for 
ages from zero to 2 Gyr, with a logarithmic time step of 0.07 between models.

Our approach then was to simplify the old populations by  only using a set
of MW globular cluster templates, and to simplify the young populations by
restricting ourselves to a single metal-rich chemical composition.  
The evolution of integrated spectra is essentially logarithmic with time;
for example, the difference between a 5 and 12 Gyr population is relatively small
compared to the large changes which occur over the first 1 Gyr. Thus our
simplification seems warranted, particularly since in this paper 
we are concerned only with identifying the younger clusters and  then studying them
in detail. The major
issue, as seen below, in separating old from young clusters, is the potential
degeneracy between very metal-poor old populations and young populations, both
of which are dominated by Balmer lines in their integrated spectra.  

Both sets of template spectra were rebinned to the same resolution and dispersion as
the M31 set, and then each template spectrum and M31 spectrum had a low
order fit subtracted. The significance of this step was that
we did not use the continuum shape to help determine the best matching
template.
A scaling factor was then determined between each template and all the available M31 
cluster spectra, and 
the reduced $\chi ^2$ calculated over
the spectral ranges of 3750-4500, 4750-5000, and 5080-5360 \AA .  These ranges were
chosen to exclude spectral regions where there are few lines, as well as where
the MW cluster spectra have no data due to bad columns in the CCD used for that set.
Specifically included were the Balmer lines from H$\beta$ to H$\zeta$, the Mg b lines, Ca II H\&K, 
and the He I lines at 4009 and 4026\AA , the last being 
prominent in OB stars and thus strong in the youngest clusters.  The noise used 
in the reduced $\chi ^2$ calculations was that
calculated from the spectra themselves.

A few logic decisions had to be made in analyzing the resultant list of
reduced $\chi ^2$ values. For most cases, the lowest reduced $\chi ^2$
occurred clearly either in the SB99 set or the MW cluster set, thus allowing
a particular cluster to be identified as young or old.  For the young clusters, the
age chosen was the average of ages where the reduced $\chi ^2$ was within 10\% of
the lowest value.   About 10\% of the
clusters were equally well fit by a MW cluster, typically a metal-poor one
with [Fe/H] $< -1$, and a metal-rich SB99 model. Nearly all of these are low S/N, and thus
the poor fits were not surprising.  Visual inspection of the spectrum clarified 
11 of these as being very young (with very blue continuum shapes), and the remaining 48 we
grouped as old.

Figure \ref{young_spectra} compares the data and best-fitting models for a range of determined ages.
We estimate the errors in determined ages for the young clusters to be about a factor
of two, which leads to M/L uncertainties of 50\%.

The model M/L values for each chosen age then allowed masses to
be estimated from the cluster integrated V band photometry, 
which is described in \ref{s:phot}. To allow a comparison of
young clusters to be made, we also calculated spectroscopic M/L values 
for the old M31 clusters from the models in \citet{leonardi}, by 
obtaining estimates of [Fe/H] for each cluster via 
Fe and Mg line indices and  assuming an age of 12 Gyr for each. The detailed
analysis of the old cluster spectra will be presented in future paper.

Many readers may be more familiar with extracting ages from diagrams that 
plot a largely age-sensitive index versus a largely metallicity-sensitive index.  To help
demonstrate the efficacy of the $\chi ^2$ approach 
we show two such diagrams. Figure \ref{indices} is a plot of a Balmer line index versus
a metal line index.  We have defined M$_{\rm avg}$=(Fe5270+Mgb)/2 
and H$_{\rm avg}$=(H$\delta _F+$ H$\gamma _F + $H$\beta$)/3, all Lick indices \citep{Worthey}.  
These indices are equivalent widths: units are \AA .  For clarity
in these diagrams, a signal-to-noise ratio cutoff was made, which eliminated 20\% of
the clusters from being plotted.
Different symbols represent MW globular clusters 
\citep [data from] []{2005ApJS..160..163S}, and four age bins for M31 clusters, which were determined
by the $\chi ^2$ method: very young ($<$ 0.1 Gyr), young (0.1 $<$ age  $<$ 1 Gyr),
intermediate (1 $<$ age $<$ 2 Gyr), and old. Clearly, there is a sequence
of M31 clusters that closely matches the sequence of MW globular clusters, and
a large spray of clusters that fall mostly in regions of higher Balmer equivalent width.
Clusters with younger ages will of course have stronger Balmer absorption, until the very
youngest ages are reached, at which point the Balmer strength declines again. 

The second diagram (figure \ref{indices2}) uses indices defined in \citet{leonardi}, namely the 
ratio of residual light in the H$\delta$ line to the nearby Fe4045 line, and
the ratio of the line at 3969\AA \ which contains both  H$\epsilon $ and CaII H, to 
the CaII K line.  The indices are unitless.  This diagram also shows the old cluster 
metallicity sequence and again distinguishes
them from the young clusters, again excepting for the extremely young clusters.

The shortcoming of using such diagrams
is that occasionally, due to the nature of real data, one index will be bad, causing 
the cluster
to look young or old in one diagram, and the opposite in another diagram.  
We are thus more confident in a fitting procedure that uses many diagnostic lines, but
are gratified that in the vast majority of cases, these two diagrams verify the ages
assigned by the  $\chi ^2$ method.

\subsection{Ages from HST/ACS color magnitude diagrams}
\citet{williams01, williams01b}  estimated ages of many young disk clusters in M31 from WFPC2 
color-magnitude diagrams (CMD) and isochrone fitting to the main sequence or
to luminous evolved stars. Four of their clusters  are bright enough to be in our spectroscopic study.
Their ages agree quite well with ours (Table \ref{ages}). Additionally, as part of
HST GO proposal 10407, we obtained ACS images of several young clusters, three of which
we report on here.

\begin{figure*}[ht]
\includegraphics[scale=0.6,clip=true]{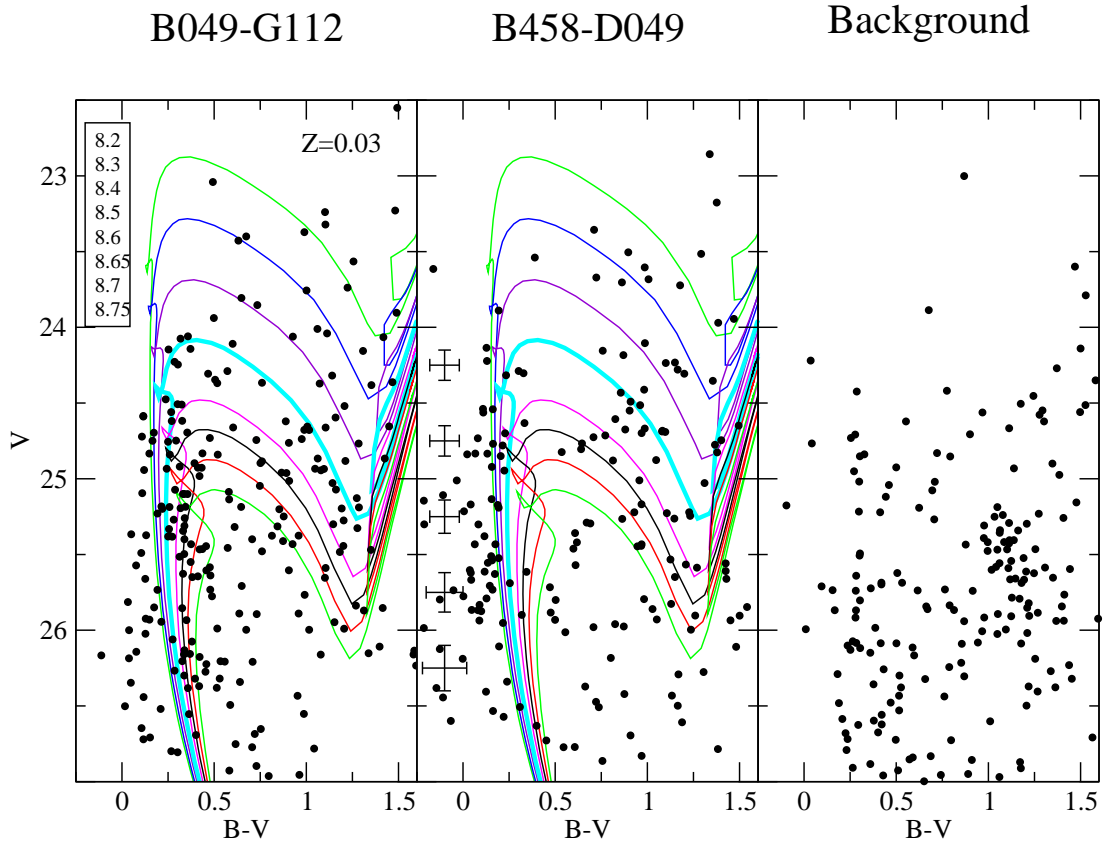}
\caption{Color magnitude diagrams obtained from HST/ACS F435W and F606W images.
Shown are two young clusters and the background field for B048-S112. Isochrones
for a range of ages at supersolar metallicity come from  \cite{cionia, cionib}. The inset
legend lists the isochrone ages shown, in log years. A distance modulus of 24.43 and reddenings of 
0.25 mag have been assumed in positioning the isochrones. Median photometric errors
at the  five indicated magnitudes are shown in the middle panel, but refer to
all three panels.  From these diagrams,
we have derived 
an age of 0.35 Gyr (log age=8.55) for B049-G112, and 0.25 Gyr (8.4) for B458-D049.
\label{cmd}}
\end{figure*}

The multidrizzle package \citep{2002hstc.conf..337K} 
was used to combine the 3 individual 
exposures taken in the F435W and F606W filters (corresponding to B and V, 
respectively).
Stellar photometry was obtained using the DAOPHOT package of
\citet{1987PASP...99..191S}, modeling the spatially variable PSFs 
for each of the  combined images separately, using only 
stars on those images.  PSFs were constructed using 5-10 bright stars which had no pixels
above a level of 20,000 counts, the point
at which an ostensible non-linearity set in and the PSF no longer
matched those of fainter stars.  Aperture corrections were also measured using these stars,
to determine any photometric offset between the psf photometry and
the aperture magnitude within 0.5\arcsec.   \citet{2005PASP..117.1049S} have
provided aperture corrections from that aperture size to infinity, 
in all ACS filters.
Generally, two passes of photometry were run. First, a star list was made and
entered into ALLSTAR, which aside of the photometry, produces a star-subtracted
image. Stars missed in the first round were located in the subtracted image 
and added to the original list.  The original frame was then measured again
by ALLSTAR.
The photometry was then placed on the standard Johnson/Kron-Cousins
VI system using the aperture corrections and synthetic transformations 
provided in \citet{2005PASP..117.1049S}. To lessen the severe problems
with crowding in these clusters, only stars that fall in an annulus with
radii  of 15 and 50 pixels (0.75 and 2.5\arcs) are shown in the
color-magnitude diagrams (figure \ref{cmd}).  The background field shown has the same area
as the cluster fields, and refers to an annulus around B049-G112 with inner
radius of 60 pixels.  Isochrones with super solar abundances 
from the Padova group \citep{cionia, cionib} have
been placed in the diagrams to allow age determination, using a distance modulus of
M31 of 24.43 and the reddenings determined above for these two clusters (0.25 for
both).  These CMDs and that of a third we have worked on (B367-G292) give ages in 
reasonable agreement with those from the spectroscopic
analysis (Table \ref{ages}). 

\subsection{Cluster integrated photometry}
\label{s:phot}
Multicolor photometry for most of the clusters in this project is
already collated in \cite{barmby}, but enough clusters are missing to
warrant remeasuring all the clusters, to allow the photometry to be
used with the spectroscopic M/L values to obtain masses.  The LGS
survey of M31 consisted of 10 separate but overlapping fields. Stellar
photometry from these fields using psf-fitting has been reported in
\cite{massey}, but the aperture photometry needed for resolved star
clusters has not yet been reported.  To limit the scope of the work,
we elected to measure objects in our catalogs only in the V band.
Targets from our entire catalog were located on the images, and
photometry for 12 separate apertures ranging from 0.7 to 16\arcs
(spaced logarithmically) was collected using DAOPHOT.  Growth curves
from these apertures were constructed, and used in an automatic
fashion to estimate the aperture which enclosed the total light of the
cluster. These apertures were then inspected on the images and
increased, if the clusters were in fact larger,  or decreased for cases
where the apertures included substantial light from objects clearly
not part of the clusters.  The local background was measured in annuli
with inner radii 1 pixel larger than the outer radius of the aperture
for the object.  The apertures used are listed in Table \ref{main}.
Extraneous stars remaining in the apertures were accounted for by
measuring their magnitudes separately, and subtracting their
contribution to the cluster aperture magnitudes.  The resultant
instrumental magnitudes were then placed on the standard V system
using stars from the \cite{massey} tables which we were measured in
the same way as the clusters.  The color term in the V mag transformation
\citep[described in][]{massey} was ignored, as it was
smaller than the errors we report. Tables \ref{main},  \ref{stars}, and
\ref{maybestars} list the results from this work.  The formal errors
in the standardized photometry were less than 0.03 mag, set by the
uncertainties in the transformations. However, since our goal was
actually cluster total magnitudes, our calculated uncertainties refer
to the uncertainty in setting the proper apertures and correcting for
extraneous objects within those apertures. Specifically, the
uncertainties were set to be equal to the difference in the magnitude
of the aperture chosen as best representing the limiting radius of the
cluster, and that of the next larger aperture in our logarithmic
spacing of aperture sizes. In practice this means that clusters in
crowded fields have larger uncertainties than those in less dense
areas. The tables also list photometry from other sources for the
objects that are outside of the LGS images; we do not list the errors
in such cases.

Comparing our aperture magnitudes with those collected from various sources 
and listed in the RBC, we find excellent
agreement over the magnitude range of 14 to 18, with an rms in the differences
of 0.18 mag in the set of 200 objects in common;  this in spite of the fact that no effort
was made to insure that the apertures used in the two data sets were the same.
Between V=18 and 20, our photometry tended to be fainter by 0.2 mag and the scatter
increased to 0.5 mag, some of which was likely due to differences in object 
identification. A further comparison was made with the 58 V magnitudes measured
for M31 clusters from archival HST images in \cite{barmbystruc}. Aside of one
cluster whose V magnitude appears to be a typo (B151-G205), the rms of the differences
of that set with the magnitudes presented here is 0.28, with no apparent systematics.

\section{The Nature of the Young Clusters}
\label{s:nature
}
\subsection{Misclassified globular clusters}
\citet{diskglobs} misclassified 15 of the young clusters as old disk
globular clusters (17\% of their disk globulars). This was pointed out
by \citet{beasley}. In some cases this was due to the low S/N of the
WYFFOS spectrum, in others the problem was misinterpretation of the
spectra.  Our new study of the clusters has resolved most of this
confusion and changed the classification of a number of clusters from
old and massive to younger, not very massive.  However, we still find
clusters with significant masses, above $10^4 $M$_\odot $, and with
ages less than 150 Myr (see Table \ref{young}).  Of the 10 clusters
with those physical characteristics, the HST or LGS images of three
confirm them as clusters similar in appearance to the populous
clusters of the LMC 
(these are B315-G038, B318-G042, and VDB0, the
latter still the most massive, young cluster known in M31, \cite{vdb}). Five appear more like OB associations, and thus may not
survive as bound clusters (B319-G044, B327-G053, B442-D033, B106D and
BH05).  The case for the other two (B040-G102, B043-G106) is not as
clear, but their LGS images are more similar to the cases like
B315-G038 than to the OB associations.

\subsection{Position}
\begin{figure*}[ht]
\includegraphics[scale=0.9,clip=true]{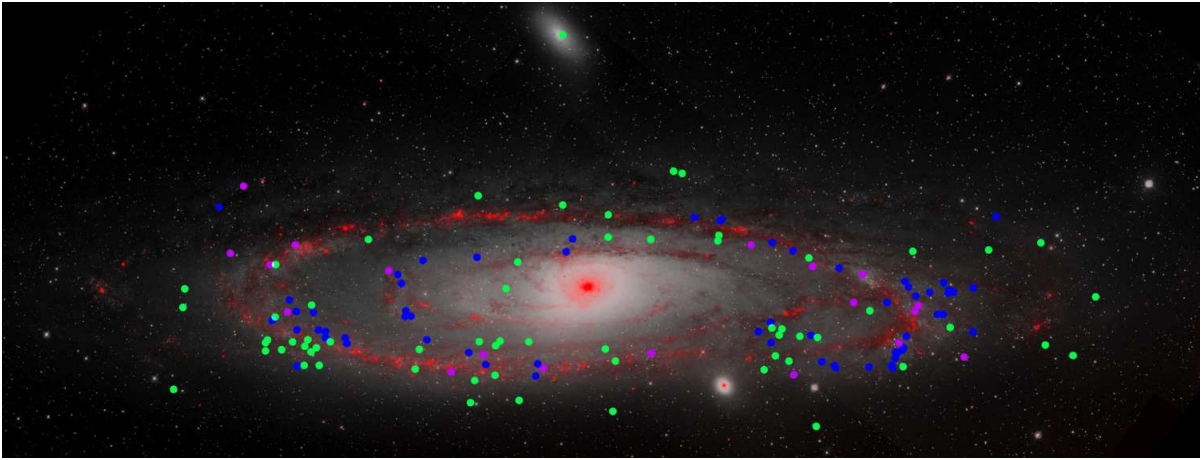} 
\caption{Spitzer/MIPS 24 micron imaging (in red on top of the optical DSS image), showing the 
location of clusters
with ages less than 0.1 Gyr as violet, between 0.1 and 0.32 Gyr as blue, and 
between 0.32 and  2 Gyr as green.
\label{mips}} 
\end{figure*}

Figure \ref{mips} shows a Spitzer/MIPS 24$\mu$ mosaic of M31
\citep{gordon} with the positions of the young clusters overlaid.
Clusters younger than 0.1 Gyr, between 0.1 and 0.32 Gyr,  
and 0.32 and 2 Gyr are shown in
different colors. The latter two groupings divide
the clusters older than 0.1 Gyr into two equal parts.  
It can be seen that the spatial distribution of the
young clusters is well correlated with the star-forming regions in
M31, with the majority associated with the 10 kpc ``ring of fire''.
The comparison of these young clusters and the warm dust emission
is distinct from the comparison of the latter with the 
location of HII regions, as we have excluded the clusters embedded
in HII regions from our sample.

\citet{gordon} and \citet{block} use the curious appearance of the
mid-IR ring - the split near the location of M32, creating the appearance
of a ``hole'' in between, and the possible offset of the ring
from the nucleus -  to suggest a  recent encounter of M32 and M31.
Both groups suggest that the split is caused by M32's passage
through the disk, and their models also produce rings offset from
M31's center, albeit
not as extreme as is observed \citep[][ produce an offset of 1$'$,
  not 6$'$]{gordon}.  An examination of our Figure \ref{mips} shows
that the cluster distribution favors the outer parts of the hole, and
is generally quite symmetric about M31's nucleus. Most models of an offset ring
assume that the inner part of the split in the observed ring is the one which should be
traced by star formation. However, this is not where most of the clusters are found.

\begin{figure}[ht]
\includegraphics[scale=0.3,clip=true]{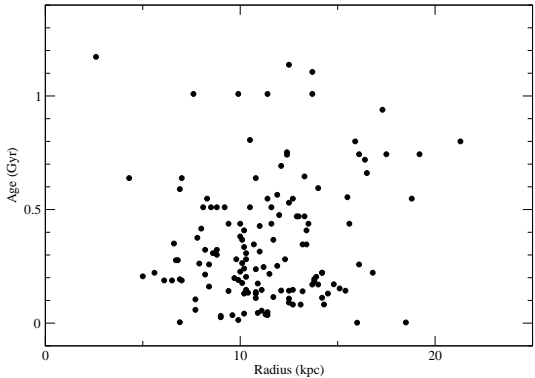} 
\caption{Ages of young clusters as a function of distance from the center
of M31, in the plane of its disk.
\label{radial_age}} 
\end{figure}

\citet{gordon} need a very recent interaction between M32 and the disk
(their model has the disk passage occurring $2 \times 10^7$
years ago) because the passage of M32 through the disk in their model
results in a burst of star formation that propagates outward through
the disk. However, we do not see any radial trends with cluster age,
which might be expected with a propagating ring of star formation (Figure \ref{radial_age}).

\citet{block} prefer a collision about $2 \times 10^8$ years ago,
which triggers expanding density waves.  Our young cluster ages range
from 0.04 Gyr to 1 Gyr, but most are between 10$^8$ and 10$^9$
years. If the ``ring of fire" was produced by a single event, as
modeled by the above authors, we might expect the age distribution of
clusters associated with it to be more peaked. The ages of the younger
clusters presented in \citet{williams01} range from around $2 \times
10^7$ to $2 \times 10^8$ years, and there is little evidence of a peak
in star formation in this age range either. These results seem to
suggest that star formation has been fairly high in this region of M31
for 1 Gyr or more.
In summary, we see no evidence of enhancement in star formation rate
or any spatial age separation, as we might expect from the M32 disk
passage.

\subsection{Kinematics}
\label{s:kin}

\begin{figure*}[ht]
\centering
\includegraphics[scale=1,clip=true]{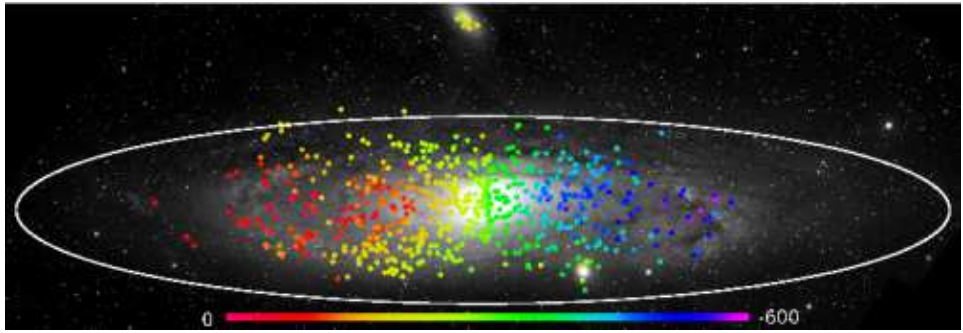}        

\caption{Disk mean velocity field, obtained from sky fibers. The color bar
  at bottom shows the velocity scale in km/s.
\label{spider_diag}} 
\end{figure*}

Do the kinematics of the young clusters bear out the disk origin
suggested by their close association (in projection) with star forming
regions in M31's disk?  M31's inner disk kinematics are more complex
than originally supposed, due to M31's bar \citep{beaton,lia}.  The
velocities from our many sky fibers, taken both as a part of regular
observing and also from entire exposures devoted to offset sky in the
crowded inner regions, give us a new way of quantifying disk rotation
throughout the inner regions where the young clusters are found. (We
plan to use these data in a study of bar kinematics, Athanassoula et
al. in preparation). Figure \ref{spider_diag} shows the disk mean
velocity field.

We also use kinematics of HII regions
that we observed as fillers in the Hectospec fields to give us an
indication of the kinematics of young disk objects. These data will be
published in Athanassoula et al. (in preparation). 

\begin{figure*}[ht]
\centering
\includegraphics[scale=0.5,clip=true]{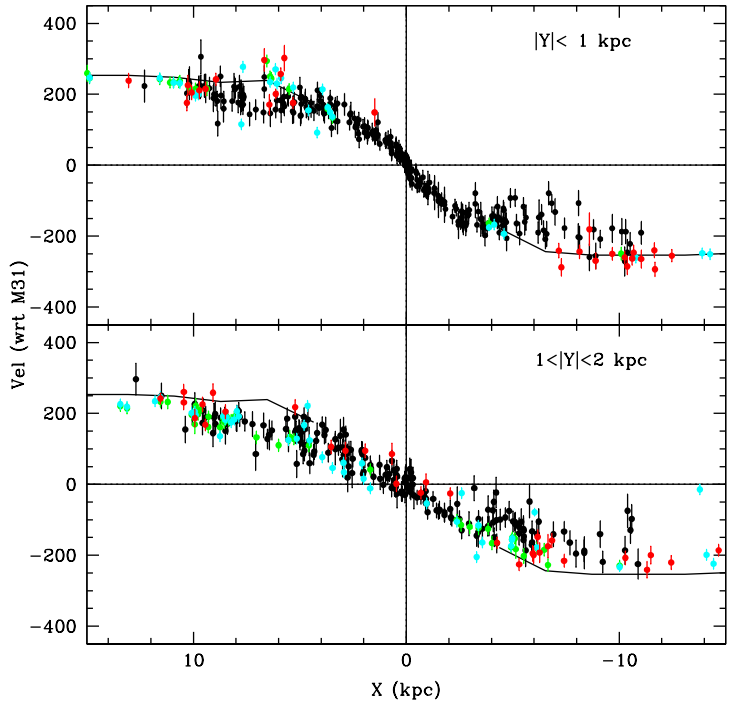}        

\caption{Plot of major axis distance X vs velocity with respect to
  M31, for objects within 1 kpc of the major axis (upper panel) and
  for objects between 1 and 2 kpc from the major axis (lower
  panel). Mean disk velocities obtained from sky fibers which show
  absorption spectra are shown in black, from sky fibers which show
  emission in green. HII region velocities are shown in cyan and
  young cluster velocities (from Table \ref{young}) in red. The
  rotation curve from \citet{kent} is shown as a solid line.
\label{dist_vel}} 
\end{figure*}

Figure \ref{dist_vel} shows the kinematics of the mean disk light, the
young clusters, and HII regions, 
versus major axis distance X, in kpc. We have split the sample into objects
which are close to the major axis ($<$ 1 kpc) and those projected from $1-2$ kpc
from the major axis, because the projection of a circular orbit looks
different in these two cases. Objects on the major axis in circular
orbits have all of their velocity projected on the line of
sight; as we move further away from the major axis, less of the
circular velocity is projected on the line of sight and so the tilt of
the distance-velocity line becomes smaller. This can be seen clearly
in Figure \ref{dist_vel} for all types of objects.

The curious flattening of the mean velocities from absorption line
spectra for major axis distances between 3 and 10 kpc is likely to be
caused by the bar. It can be seen that the young clusters follow the
disk mean velocity curve from absorption spectra quite well, and show
an even better correlation with the kinematics of the youngest
objects: HII regions and sky fibers showing emission spectra.

This kinematic analysis confirms our spatial association of the young
clusters with the star forming young disk in M31.

\subsection{Masses of the clusters}
\label{s:masses}

The M/L values obtained from the spectroscopic age estimates can be combined
with the V band photometry  to derive masses of all the observed M31 clusters, young
and old. Reddening values are of course also needed, and a large number
of E(B--V) values were derived from photometry in \cite{barmby}.  They and we consider only
the total reddenings, foreground and internal to M31. The methodology
used in  \cite{barmby} meant that the reddenings would only be valid for old clusters,
and indeed few of the clusters we have identified here as young were included in 
their study, thus reddenings for those objects are needed.    
Therefore, we elected to rederive the reddenings for all of the clusters
in our study, young and old.  

In the case of the young
clusters  we compared the fluxed spectra with the SB99 model spectra of the appropriate age.
As described above, the ages were obtained by matching spectral line features in the
observed and model spectra, not by comparing the continua shapes.  Once the ages
have been found in this way, differences in the continua shapes may be assumed to
be due to reddening, except for a few cases where a late-type star, whether member
or not,  clearly dominates the redder wavelengths as evidenced by the presence of
TiO bands. For those cases, we use the mean reddening for the young clusters of 0.28.

For the old clusters we did not use models, but rather the
sample of spectra themselves.  Initial values
of reddenings were obtained from \cite{barmby}, and were used to deredden the spectra of
those clusters with E(B$-$V)$< 0.4$, about 190 in number (there are about 350 old clusters in our 
spectroscopic sample).
These spectra were ordered in metallicity, which was estimated from the spectral line indices 
as mentioned above, rebinned to a coarse grid in wavelength, and normalized to 
have the same intensity at the arbitrary wavelength of 5000\AA .  
Interpolation formulae were developed from these spectra,  via a least squares method to
avoid bad spectra, 
for intensity as a function of both wavelength and
metallicity. As a result,  a cluster spectrum of arbitrary
metallicity could be created, dereddened to the accuracy of the \cite{barmby} 
reddenings.  The
individual spectra in this low reddening sample were then compared with the appropriate interpolated spectra, 
and reddenings were adjusted as needed to bring their continua shapes
closer to that of the expected template shape.  The method is thus similar to
methods that use the metal abundance to predict the intrinsic broad band colors,
and then require the derived reddening to reproduce the observed colors.

The overall goal in working with the low reddening sample was 
to retain the mean value of the
reddenings found in  \cite{barmby}, but to correct those that varied significantly. After cleaning
up those reddenings,  the interpolation formulae were then used 
to derive reddenings for the 150 clusters for which we have
spectra and whose reddenings were not measured in \cite{barmby}.  Thus while we have
not improved upon the absolute levels of the M31 old cluster reddenings, we believe we have improved
the precision of the values in a relative sense, and as well have nearly doubled the number
of reddenings available.  About 10\% of the spectra were taken during nights when the ADC was
not operating properly, thus we can't use the continuum shape to estimate reddenings.  For
objects whose only spectra were taken on those nights, we assume the average reddening of 0.28.

A comparison of our derived E(B$-$V) values (which range up to 1.4 mag) 
and those in  \cite{barmby} results in 
a scatter of 0.17 mag rms, which is good enough for our overall goal of comparing
the M31 cluster system in bulk with that of other galaxies.  Interestingly, both the
young cluster and old cluster groups have clusters with  E(B$-$V)$ >  0.5$, though the
highest measured value  ( E(B$-$V)=1.4) is still found in the old cluster B037-V327, probably a selection 
effect since that cluster also has the highest luminosity in all of M31. The young cluster
reddenings are listed in Table \ref{young}; those of the old clusters will be
presented in a subsequent paper.   By using the position of blue-plume stars in 
the color-magnitude diagram, \cite{massey2} estimated the average reddening
for young stars in M31 to be 0.13 mag, significantly lower than the mean of the
clusters younger than 100 Myr presented here, which may place a constraint
on the accuracy of the values presented here.

\begin{figure}
\centering
\includegraphics[scale=0.4,clip=true]{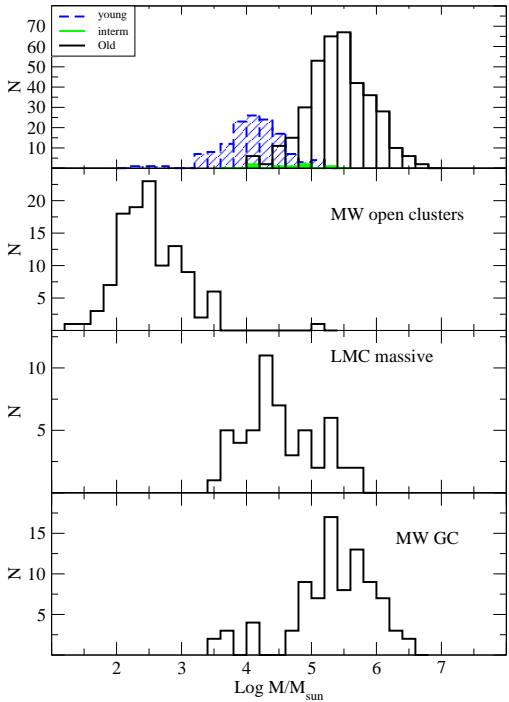}
\caption{Mass histograms, from top to bottom, M31 clusters (young, intermediate and old), 
Milky Way open clusters, LMC massive clusters and Milky Way globular clusters. The young M31clusters are shown in a hatched histogram, the intermediate as solid, and the old as open.}
\label{massdist}
\end{figure}

The mass histogram for the all of the young clusters is shown in
Figure \ref{massdist}.  We have also shown the mass distribution of
Milky Way open clusters within 600 pc of the Sun. This is based on the
sample of \cite{khar}, with mass calculations by \cite{henny}.  The
Kharchenko catalog is the most homogeneous and complete
catalog of open clusters in the solar neighborhood currently
available, and is based on a stellar catalog complete to V=11.5. The
cluster masses were estimated by counting the number of cluster members
brighter than the limiting magnitude, then correcting for the stars
fainter than this using a Salpeter mass function and a lower mass
limit of 0.15 M$_\odot$.  This catalog does not include the most
massive clusters in the Galaxy because of its relatively small sample
size; for example, there have been recent discoveries of more distant
young clusters which may have masses as high as $10^5 $M$_\odot$
(e.g. \cite{clark}), and we add Westerlund 1 to the histogram as an example.  The
Milky Way globular and LMC young massive cluster histograms are shown
in the bottom two panels  \citep[from][]{dean}.  These mass estimates are
based on King model fits.

Obviously, M31 clusters with masses less than $\approx 10^3 $M$_\odot
$ and ages greater than a few $\times 10^7$ years are too faint to be
part of this study, and await a future study.  \cite{krienke} estimate
over 10000 such clusters in the disk of M31; these would form the low
mass tail in the mass distribution of Figure \ref{massdist}.

Nonetheless, there is a trend in cluster mass,
with the Milky Way open clusters having the lowest median mass, the
Milky Way and M31 globulars the highest, and the LMC young massive clusters
and the M31 young clusters in between. This trend is consistent with a
single cluster IMF plus disruption, taking into account the small size
of the volume searched for clusters in the Milky Way.

\subsection{Cluster survival}
\label{s:survival}
Would we expect these young M31 clusters to survive as they age, or to
disrupt? One of the main processes that leads to cluster disruption is
2-body relaxation enhanced by an external tidal field
\citep{spitzerharm}.  The lower-mass clusters suffer more strongly
from  relaxation effects. Another property of the cluster itself
which will affect its survival is its density --- lower-density
clusters will disrupt more quickly \citep{spitzer58}. Thus we would
expect that massive, concentrated clusters such as B018 and BH05 would
be more likely to survive.

\begin{figure}[ht]
\centering
\includegraphics[scale=0.4,clip=true]{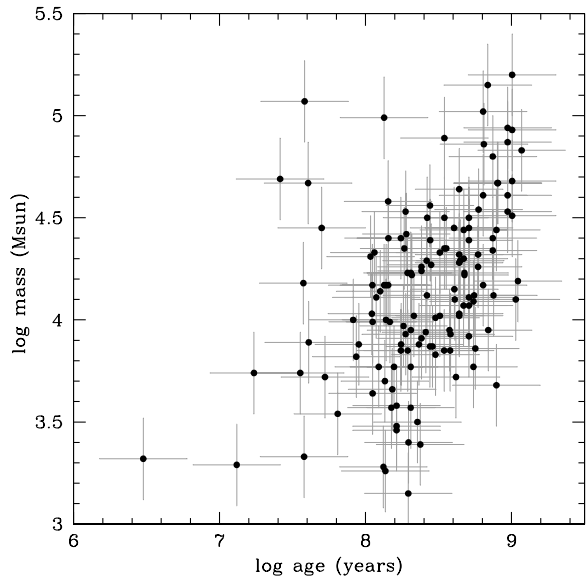}

\caption{Age-mass diagram for our young and intermediate-age clusters.
\label{agemass}}
\end{figure}

\citet{boutloukos} derive an empirical expression for the disruption
of clusters as a function of their mass, studying cluster populations
in the solar neighborhood, the SMC, M33 and M51.  \citet{whitmore}
point out that observational selection effects could mimic the
decrease in the number of clusters with age which Boutloukos et
al. ascribe to cluster disruption. However, this is almost certainly
not true of the solar neighborhood open clusters studied by
\citet{henny} using a similar analysis.  We show the age-mass diagram
for the young M31 clusters in Figure \ref{agemass}. While our sample
is clearly very incomplete below $10^8$ years, the diagram shows some
similarity to the LMC cluster age-mass diagram of \citet{degrijs} in
the age range we cover. Unfortunately, we do not expect our catalog to
be complete enough to permit an analysis using the techniques of
Boutloukos et al.

Environmental effects also control the tidal stripping of the
cluster. For stars whose orbits are mostly confined to the disk,
encounters with giant molecular clouds and spiral arms contribute to
their disruption \citep{spitzer58,gieles07}. For clusters whose orbits
are not confined to the disk, bulge and disk shocking are more
important \citep{ostriker72,aguilar}. The similarity of the M31 young
cluster kinematics to that of other young disk objects suggests
strongly that these clusters are confined to M31's disk plane,
so giant molecular clouds should be the relevant external
disruptor. \citet{gieles08} show that disruption times for clusters in
galaxies ranging in size from M51 to the SMC, scale with molecular gas density
in the expected way. M31's molecular gas density is highest near the
''ring of fire" where many of our clusters are found
\citep{loinard99}. This density is similar to the molecular gas
density in the solar neighborhood \citep{dame93}. Thus we would expect
the survival due to giant molecular cloud interactions of the M31
young clusters to be similar to that of the solar neighborhood open
clusters.

We expect that most of these young clusters will be
disrupted in the next Gyr or so \citep[][derive a disruption time of 1.3 Gyr
for a cluster of mass 10$^4 M_\odot$ in the solar
neighborhood]{henny}. However, some of the more massive and
concentrated of the young clusters will likely survive for longer.

\section{Summary}

We present a new catalog of 670 M31 clusters, with
accurate coordinates. In this paper we focus on the 140 clusters
(many originally classified in the literature as globular clusters) which
have ages less than 2 Gyr: most have ages between $10^8$ and $10^9$
years. Using high-quality MMT/Hectospec spectra, excellent ground
based images, and in some cases, HST images, we explore the nature of
these clusters.  With the exception of NGC 205's young cluster,
they have spatial and kinematical properties consistent with formation
in the star-forming disk of M31. Many are located close to the 10 kpc
``ring of fire'' which shows active star formation. 
The age distribution of our clusters, plus that of the younger
clusters of \citet{williams01}, shows no evidence for a peak in star
formation there between $2 \times 10^7$ and $10^9$ years ago, which we
might expect if the ring was created by a recent passage of M32
through the disk, as suggested by \citet{gordon} and \citet{block}.

We have estimated their masses using spectroscopic ages and M/L
ratios, (in some cases) ACS color-magnitude diagrams, and new
photometry from the Local Group Survey. The clusters have masses
ranging from 250 to 150,000 $M_\odot$. These reach to higher values
than the known Milky Way open clusters, but it must be remembered that
our sample of open clusters in the Milky Way is far from complete. The
most massive of our young clusters overlap the mass distributions of
M31's old clusters and the Milky Way globulars.

Interestingly, although most of the young clusters show the
low-concentration structure typical of the Milky Way open clusters, a
few have the high concentrations typical of
the Milky Way globulars and the old M31 clusters. We estimate that
most of these young clusters will disrupt in $1-2$ Gyr, but the
massive, concentrated clusters may well survive longer.

\acknowledgments
We would like to thank Dan Fabricant for leading the effort to
design \& build the Hectospec fiber positioner and spectrograph,
Perry Berlind \&  Mike Calkins for help with
the observations, John Roll, Maureen Conroy \&  Bill Joye for their many 
contributions to the Hectospec software development project, and  Phil Massey,   
Pauline Barmby  \&  Jay Strader for comments and data tables on M31.
HLM was supported by NSF grant AST-0607518, and 
would like to thank Dean McLaughlin for helpful conversations.
Work on this project has also been supported by HST grant GO10407.

\clearpage



\LongTables




\clearpage
\pagestyle{empty}
\begin{deluxetable}{llllllrll}
\tablecolumns{8}
\tablewidth{0pc}
\tablecaption{Cleaned Cluster Catalog \label{main}} 

\tablehead{\colhead{Object} & \colhead{RA}  &\colhead{Dec} &\colhead{V}  &\colhead{type}   &\colhead{S\tablenotemark{a}}  &\colhead{Ap} &\colhead{P\tablenotemark{b}} &\colhead{C\tablenotemark{c}} \\
\colhead{} & \multicolumn{2}{c}{2000} &\colhead{}    &\colhead{} &\colhead{} &\colhead{\arcs} &\colhead{} &\colhead{}
}
\startdata
{\it SH01}&	0:32:41.44&	40:01:41.4&	15.82&	&	&		&	G&		\\
G001-MII&	0:32:46.53&	39:34:40.6&	13.75&	old&	B&		&	B&	H	\\
G002-MIII&	0:33:33.77&	39:31:18.9&	15.81&	old&	B&		&	B&		\\
B290&	0:34:20.94&	41:28:18.1&	17.14&	old&	P&		&	B&		\\
{\it BA21}&	0:34:53.83&	39:49:40.5&	16.64&	&	&		&	G&		\\
{\it B412}&	0:34:55.28&	41:32:26.4&	17.36&	&	&		&	G&		\\
{\it B413}&	0:35:13.00&	41:29:07.8&	18.12&	&	&		&	G&		\\
{\it BA22}&	0:35:13.60&	39:45:37.1&	&	&	&		&	&		\\
\enddata
\tablenotetext{a}{Source of velocity: HS=this paper; B=\citet{barmby}; P=\citet{perrett}}
\tablenotetext{b}{Source of photometry: L=this paper; B=\citet{barmby}; G=\citet{galleti2}, H=\citet{huxor}}
\tablenotetext{c}{Source of classification as a cluster: S=spectrum from this paper indicates a cluster; L=LGS image indicates non-stellar; H=HST image indicates a cluster; objects with blank entries in this column and in the velocity source column should still be considered ``candidates''}
\tablenotetext{d}{not a cluster in \citet{barmby}}
\tablenotetext{e}{not a cluster in \citet{crampton85}}
\tablenotetext{f}{not a cluster in \citet{racine}}
\tablenotetext{g}{\citet{grillmair} probably observed a different object}
\end{deluxetable}

\clearpage
\pagestyle{empty}
\begin{deluxetable}{lrrcccl}
\tablecolumns{7}
\tablewidth{0pc}
\tablecaption{Young Clusters, Ordered by Mass \label{young}}

\tablehead{\colhead{Object}  &\colhead{\footnotesize{E(B$-$V)}}  &\colhead{Vel}  &\colhead{Log Age } &\colhead{Log Mass} &\colhead{$\chi^2$} &\colhead{Comments} \\
\colhead{}  &\colhead{} & \colhead{\kms}  &\colhead{yr} &\colhead{M$_{\sun}$}   &\colhead{}
}
\startdata
B316-G040&                         	0.20&   	-359.1$\pm	15$& 	9.0&	5.2& 	0.7& 	           	\\ 
B392-G329&                         	0.25&   	-72.8$\pm 	16$& 	8.9&	5.2& 	0.7& 	           	\\ 
B324-G051\tablenotemark{f.i}&      	0.28&   	-218.2$\pm	15$& 	8.8&	5.1& 	1.4& 	NGC205-HubV	\\ 
VDB0-B195D\tablenotemark{b}&       	0.28&   	-551.8$\pm	20$& 	7.6&	5.1& 	2.7& 	emiss      	\\ 
B315-G038\tablenotemark{e,g,h,j}&  	0.35&   	-555.2$\pm	17$& 	8.2&	5.0& 	1.1& 	           	\\ 
B222-G277\tablenotemark{g,i}&      	0.28&   	-287.4$\pm	12$& 	8.9&	5.0& 	0.9& 	           	\\ 
B307-G030\tablenotemark{g}&        	0.25&   	-454.6$\pm	20$& 	9.0&	4.9& 	0.4& 	           	\\ 
DAO30&                             	0.25&   	-504.4$\pm	18$& 	9.0&	4.9& 	0.5& 	           	\\
\enddata
\tablenotetext{a}{Not a cluster in \citet{cohen}}
\tablenotetext{b}{Star in field}
\tablenotetext{c}{Noted as young in \citet{vdb}}
\tablenotetext{d}{Noted as young in \citet{hodge79}}
\tablenotetext{e}{Noted as young in \citet{elsonw}}
\tablenotetext{f}{Noted as young in \citet{barmby}}
\tablenotetext{g}{Noted as young in \citet{fusipecci}}
\tablenotetext{h}{Noted as young in \citet{burstein}}
\tablenotetext{i}{Noted as young in \citet{beasley}}
\tablenotetext{j}{Noted as young in \citet{williams01}}
\end{deluxetable}

 \clearpage
\pagestyle{empty}
\begin{deluxetable}{lllrlll}
\tablecolumns{7}
\tablewidth{0pc}
\tablecaption{Stars \label{stars}}

\tablehead{\colhead{Object\tablenotemark{a}} & \colhead{RA}  &\colhead{Dec} &\colhead{Vel\tablenotemark{b}}   &\colhead{V}   &\colhead{P\tablenotemark{c}}  &\colhead{C\tablenotemark{d}}  \\
\colhead{} & \multicolumn{2}{c}{2000} &\colhead{\kms}  & \colhead{} & \colhead{}  &\colhead{} 
}
\startdata
B133D&	0:34:10.99&	39:50:50.2&		&	18.18&	G&	B	\\
BH01&	0:34:11.92&	39:24:11.6&		&	&	&	H	\\
B135D&	0:35:32.39&	41:19:47.5&		&	17.76&	G&	B	\\
DAO3&	0:35:51.90&	40:20:55.7&	 -83.8$\pm	 36$&	&	&	S	\\
B137D&	0:36:20.90&	40:56:00.0&		&	18.51&	G&	B	\\
B145D-G013&	0:36:38.76&	42:16:28.5&	 -84.6$\pm	 20$&	17.53&	G&	S	\\
B146D&	0:36:42.42&	41:38:19.9&		&	18.07&	G&	B	\\
B153D-G017&	0:37:12.55&	39:50:16.8&	  -4.6$\pm	 24$&	16.66$\pm0.04$&	L&	SL	\\
\enddata
\tablenotetext{a}{The suffix ``x'' indicates that coords correspond to a real object, but which was wrongly identified in the initial input catalog.}
\tablenotetext{b}{All velocities from Hectospec}
\tablenotetext{c}{Source of photometry: L=this paper; B=\citet{barmby}; G=\citet{galleti2}}
\tablenotetext{d}{Source of classification as a star: S=spectrum from this paper indicates a star; L=LGS image indicates stellar FWHM; H=HST image indicates a star; B=\citet{barmby} indicated a star}
\tablenotetext{e}{NB61 is an M star,  the velocity of -646 listed  \citet{galleti2}
was likely the result of a velocity template mismatch.}
\tablenotetext{f}{Multiple stars or asterisms in M31.}
\tablenotetext{g}{The velocity of this F supergiant star indicates is it probably a member of
the M31 giant stream. Details to be presented in a subsequent paper.}

\end{deluxetable}

\clearpage
\pagestyle{empty}
\begin{deluxetable}{lllrlrll}
\tablecolumns{8}
\tablewidth{0pc}
\tablecaption{Possible Stars \label{maybestars}}

\tablehead{\colhead{Object\tablenotemark{a}} & \colhead{RA}  &\colhead{Dec} &\colhead{Vel}   &\colhead{V}    &\colhead{S\tablenotemark{b}}  &\colhead{P\tablenotemark{c}} &\colhead{C\tablenotemark{d}}\\
\colhead{} & \multicolumn{2}{c}{2000} &\colhead{\kms}  &\colhead{}  &\colhead{}  &\colhead{}  &\colhead{}
}
\startdata
SK015B&	0:37:09.77&	40:11:36.7&		&	16.15$\pm0.07$&	&	L&	L	\\
SK024B&	0:38:11.02&	40:38:52.3&		&	19.13$\pm0.08$&	&	L&	L	\\
B172D&	0:38:42.20&	40:12:12.0&	 -90.4$\pm	 22$&	17.72$\pm0.04$&	HS&	L&	SL	\\
SK059C&	0:38:53.74&	40:38:30.4&		&	18.50$\pm0.04$&	&	L&	L	\\
B177D&	0:38:56.19&	40:03:14.9&	-150.6$\pm	 21$&	17.66$\pm0.07$&	HS&	L&	SL	\\
SK068C&	0:39:23.47&	40:14:40.4&		&	16.93$\pm0.06$&	&	L&	L	\\
B183D&	0:39:42.29&	40:40:32.0&		&	18.14$\pm0.03$&	&	L&	L	\\
SK034B&	0:39:51.23&	40:45:33.1&		&	19.60$\pm0.09$&	&	L&	L	\\
\enddata
\tablenotetext{a}{The suffix ``x'' indicates that coords correspond to a real object, but which was wrongly identified in the initial input catalog.}
\tablenotetext{b}{Source of velocity: HS=this paper; B=\citet{barmby}; K=\citet{kim}; P=\citet{perrett}}
\tablenotetext{c}{Source of photometry: L=this paper; B=\citet{barmby}; G=\citet{galleti2}}
\tablenotetext{d}{Source of classification as a star: S=spectrum from this paper exists but is inconclusive; L=LGS image indicates stellar FWHM; B=\citet{barmby} indicated a star}
\tablenotetext{e}{The colors of SK106A indicate it is an M star,  the velocity of -890 reported in \citet{kim}
was likely the result of a velocity template mismatch  \citep[see also][]{lee} }
\end{deluxetable}

\clearpage
\pagestyle{empty}
\begin{deluxetable}{lllrr}
\tablecolumns{4}
\tablewidth{0pc}
\tablecaption{Galaxies \label{gals}}

\tablehead{\colhead{Object} & \colhead{RA}  &\colhead{Dec} &\colhead{Z\tablenotemark{a}}   \\
\colhead{} & \multicolumn{2}{c}{2000} &\colhead{} 
}
\startdata
G006&	0:35:04.77&	42:35:32.2&	0.06568	\\
G007&	0:35:13.70&	42:35:18.5&	0.08222	\\
DAO1&	0:35:18.89&	40:09:11.1&	0.20722	\\
B136D-G008&	0:35:52.43&	39:57:08.0&	0.10042	\\
B414&	 0:36:19.3&	42:16:30.9&	0.08603	\\
B294-G012&	0:36:32.94&	40:24:53.4&	0.03588	\\
B143D-D005&	0:36:35.68&	40:28:15.2&	0.09974	\\
B296-G015&	0:36:46.88&	40:30:47.1&	0.10004	\\
\enddata
\tablenotetext{a}{All redshifts from Hectospec. Median velocity error is 17\kms. Entries without redshifts were classified background
based on images.}
\tablenotetext{b}{\citet{perrett} had velocity indicating M31 membership}

\end{deluxetable}

\clearpage
\pagestyle{empty}
\begin{deluxetable}{lll}
\tablecolumns{3}
\tablewidth{0pc}
\tablecaption{Missing objects or bad coordinates \label{nobodyhome}} 
\tablehead{\colhead{Object} & \colhead{RA}  &\colhead{Dec}   \\
\colhead{} & \multicolumn{2}{c}{2000}    
}
\startdata
SH05&     	0:38:55.20&	41:10:28.0	\\ 
SH06&     	0:39:19.30&	41:10:29.0	\\ 
SH08&     	0:40:31.40&	40:26:27.0	\\ 
B450&     	0:40:46.8& 	41:40:27.0	\\ 
NB60&     	0:42:26.68&	41:18:10.7	\\ 
V229&     	0:42:34.50&	40:55:44.0	\\ 
NB57&     	0:42:36.13&	41:13:13.2	\\ 
NB44&     	0:42:46.52&	41:17:58.4	\\ 
NB42&     	0:42:46.97&	41:17:35.2	\\ 
B353-CFA1&	0:42:47.41&	41:15:40.5	\\ 
NB104&    	0:42:56.22&	41:14:14.8	\\ 
SH14&     	0:44:14.8& 	41:55:23.9	\\ 
M060&     	0:45:06.83&	41:38:57.8	\\ 
DAO93&    	0:47:46.20&	41:44:55.0	\\ 
SH25&     	0:51:58.70&	41:35:17.0	\\ 

\enddata
\end{deluxetable}

\clearpage
\pagestyle{empty}
\begin{deluxetable}{llll}
\tablecolumns{4}
\tablewidth{0pc}
\tablecaption{Comparison of ages \label{ages} }
\tablehead{\colhead{Object} & \colhead{Age (Gyr)}  &\colhead{Age (Gyr)}  &\colhead{ref}  \\
\colhead{} & \colhead{CMD} &\colhead{Spec} &  \colhead{}
}
\startdata
B049-G112&   	0.35 &	0.44 &  this paper\\ 
B315-G038&   	0.10 &	0.13 & \cite{williams01} \\ 
B318-G042&   	0.056 &	0.12 & \cite{williams01b}  \\ 
B319-G044&   	0.10 &	0.28 & \cite{williams01} \\ 
B342-G094&   	0.16 &	0.15 & \cite{williams01}  \\ 
B367-G292&   	0.20 &	0.20 & this paper \\ 
B458-D049&   	0.25 &	0.51 & this paper\\ 

\enddata
\end{deluxetable}

\end{document}